\documentclass[aps, prb, twocolumn, superscriptaddress]{revtex4-2}

\usepackage{amsmath,amssymb,bm}
\usepackage[dvipdfmx]{graphicx}
\usepackage{dcolumn}
\usepackage{color}
\usepackage{accents}
\usepackage{mathtools}
\usepackage{ulem}
\usepackage{url}
\usepackage[dvipdfmx]{hyperref}

\begin{document}

\title{Spin-triplet superconductivity induced by Zeeman fields}
\date{\today}

\author{Kota Tabata}
\affiliation{Department of Applied Physics, Hokkaido University, Sapporo 060-8628, Japan}
\author{Tomoya Sano}
\affiliation{Department of Applied Physics, Hokkaido University, Sapporo 060-8628, Japan}
\author{Satoshi Ikegaya}
\affiliation{Department of Applied Physics, Hokkaido University, Sapporo 060-8628, Japan}
\author{Yasuhiro Asano}
\affiliation{Department of Applied Physics, Hokkaido University, Sapporo 060-8628, Japan}

\begin{abstract}
We study the stability of spin-triplet superconductivity in the presence of Zeeman fields and spin-orbit interactions.
The magnetically active potentials induce the subdominant pairing correlations depending on 
the relative orientations among the three vectors in spin space: a $\bm{d}$ vector of 
a spin-triplet superconducting state, a Zeeman potential $\bm{V}$, 
and a potential representing spin-orbit interaction $\bm{\alpha}$. 
Although the induced pairing correlations do not form the pair potential, they affect 
the stability of the superconducting states. 
When the three vectors form a nonzero scalar product $\bm{\alpha}\times\bm{V}\cdot \bm{d}$, 
 superconductivity appears at high Zeeman fields.     
We show that the spin-singlet even-parity Cooper pairs stabilize such Zeeman-field induced superconductivity.  
The specific heat, the magnetization and the magnetic susceptibility are calculated for the Zeeman-field induced 
superconducting phase.
\end{abstract}

\maketitle

\section{Introduction}
The pair potentials in superconductors (SCs) are usually classified into two 
symmetry classes: spin-singlet even-parity and spin-triplet odd-parity.
Most of SCs discovered to date, such as monoelement metals and high-$T_c$ cuprates, belong to 
 spin-singlet symmetry class. In contrast, spin-triplet superconductors are very rare.
The topological-material-based compound $\mathrm{Cu}_x\mathrm{Bi_2Se_3}$ and several uranium compounds, such as 
$\mathrm{UPt_3}$, $\mathrm{UBe_{13}}$, $\mathrm{UGe_2}$ and $\mathrm{UTe_2}$, are candidates for spin-triplet superconductors~\cite{stewart:prl1984, ott:prl1983, saxena:nature2000, ran:science2019, jiao:nature2020, hor:prl2010, sasaki:prl2011}. 
The discovery of spin-triplet superconductors is still a challenging issue in physics of superconductivity. 
 In order to confirm spin-triplet superconductivity in newly discovered SCs, 
 it is necessary to observe phenomena unique to spin-triplet superconductivity 
characterized by a $\bm{d}$ vector. 
The unchanged Knight shift across the transition temperature $T_c$ is such a characteristic response.
While a Zeeman potential $\bm{V}=\mu_{\mathrm{B}} \bm{H}$ always suppresses 
spin-singlet superconductivity, it leaves $T_c$ unchanged in a spin-triplet SC with $\bm{d} \perp \bm{V}$.  
Moreover, several layered SCs indicate the stable superconducting phase in high Zeeman fields 
parallel to the layers~\cite{meul:prl1984,uji:nature2001,maryenko:sciadv2026}.
Such Zeeman-field induced superconductivity (ZFIS) can be considered as an 
evidence of spin-triplet SCs. 
Theoretically only Jaccarino-Peter compensation~\cite{jaccarino:prl1962} is known as a 
widely-accepted mechanism for the ZFIS.
In this paper, we provide an alternative scenario for the phenomenon.

In layered SCs,  magnetic fields parallel to the layer tell us 
clear signals of a $\bm{d}$ vector because orbital effects are negligible.
However, the magnetic response of a layered SC often depends also on a vector $\bm{\alpha}$ 
that represents spin-orbit interaction (SOI). 
For example in spin-singlet SCs,
the critical Zeeman field goes beyond 
the Pauli limit 
$H_{\mathrm{P}}$
~\cite{maki_tsuneto:ptp1964,chandrasekhar:apl1962,clogston:prl1962}  
for $\bm{V} \perp \bm{\alpha}$~\cite{klemm:prb1975,tedrow:prb1982,agosta:prb2012,lee:prb2000,bauer:prl2004,kimura:prl2007,hoshi:sr2022,foner:pra1973,prober:prb1980,saito:natphs2016,uher:prb1986,sano:prb2026}.  
The phenomenon is interpreted as if the spin-orbit interactions weaken the 
pair-breaking effect by a Zeeman field. 
For spin-triplet SCs, the situation would be more complicated because 
the stability of superconducting states would depend on the relative configurations among 
$\bm{d}$, $\bm{V}$ and $\bm{\alpha}$ in spin space.

In this paper, 
we first discuss the stability of superconducting states in typical five 
configurations of $\bm{d}$, $\bm{V}$ and $\bm{\alpha}$
 by using a two-valley model in which 
 $\bm{\alpha}$ and $\bm{d}$ are independent of wavenumber $\bm{k}$~\cite{sano:prb2026}.
It is possible to derive the analytic expressions of gap equation and the superfluid weights of 
induced pairing correlations, which enables us to understand the physics behind the 
stability or instability of superconducting states.
To calculate $T_c$, we solve the Gor'kov equation for the Bogoliubov-de Gennes Hamiltonian
and find the free-energy minimum simultaneously.
The instability of a uniform superconducting state is characterized well by
the odd-frequency Cooper pairing correlations because they 
always decrease the superfluid weight~\cite{asano:prb2015}.
The SOI $\bm{\alpha} \parallel \bm{d}$ and the Zeeman field $\bm{V} \perp \bm{d}$ do not affect the
stability of superconducting states, which is described by the absence of the 
odd-frequency pairing correlations.
On the contrary, 
the SOI $\bm{\alpha} \perp \bm{d}$~\cite{frigeri:njp2004} and the Zeeman field $\bm{V} \parallel \bm{d}$ 
make the superconducting state unstable. In other words, such magnetic potentials induce 
odd-frequency Cooper pairs as a subdominant pairing correlation.
In particular for $\bm{\alpha}\times\bm{V}\cdot \bm{d} \neq 0$,  
induced spin-singlet even-frequency Cooper pairs support spin-triplet superconductivity in high Zeeman fields, 
which explains a mechanism of ZFIS~\cite{sano:arXiv2602}. 
Secondly, we discuss the thermal and the magnetic properties 
of the Zeeman-field-induced superconducting phase by calculating specific heat, magnetization, 
and magnetic susceptibility. 
 
This paper is organized as follows.
We explain our theoretical model in Sec.~\ref{sec:model}. We also explain the reasons we analyze the superfluid weights.
The results of the anomalous Green's function, the temperature dependence of the pair potential, and 
the phase diagram in Zeeman fields are presented in Sec.~\ref{sec:results}.
The physical properties of Zeeman-field induced superconducting phase are shown in Sec.~\ref{sec:property}.
The conclusion is given in Sec.~\ref{sec:conclusion}.
We use units of $k_B=\hbar=c=1$, where $k_B$ is the Boltzmann constant and $c$ is the speed of light.

\section{Two-valley Model}\label{sec:model}
We investigate the stability of spin-triplet superconductivity 
under magnetically active potentials by using a two-valley model 
for two-dimensional electronic states. 
The advantage of the two-valley model and the connection of the model to real materials 
are discussed later.
The two-valleys are represented by two Fermi surfaces that surround certain points
of $\bm{K}$ and $\bm{K}'$ or $-\bm{K}$ in momentum space.
The normal state Hamiltonian reads
\begin{align}
    \check{H}_{\mathrm{N}} (\bm{k})
    &=
    \hat{\xi}_{\bm{k}} \hat{\sigma}_0 - \bm{V} \cdot \hat{\bm{\sigma}}\hat{\rho}_{0}
    - \bm{\alpha} \cdot \hat{\bm{\sigma}}\hat{\rho}_{z},\\
    \hat{\xi}_{\bm{k}} =& \frac{1}{2m}(\bm{k}\hat{\rho}_{0}-\bm{K}\hat{\rho}_{z})^{2} - \mu,
\end{align}
with $\bm{V}=\mu_{\textrm{B}}\bm{H}$, where $\mu$ is the chemical potential, $\mu_{\textrm{B}}$ is the Bohr magneton, 
$\bm{H}$ represents an external Zeeman field applied parallel to the two-dimensional plane.
The Pauli's matrices in spin and valley spaces are denoted by $\hat{\bm{\sigma}} = (\hat{\sigma}_x, 
\hat{\sigma}_y, \hat{\sigma}_z)$ and $\hat{\bm{\rho}} = (\hat{\rho}_x, 
\hat{\rho}_y, \hat{\rho}_z)$.
The unit matrices in the two spaces are denoted by $\hat{\sigma}_{0}$ and $\hat{\rho}_{0}$.
Although very similar electronic structures are realized in transition-metal 
dicalcogenides~\cite{saito:natphs2016,xi:natphs2016,barrera:natcom2018}, we do not focus on 
any specific materials in this paper.
The antisymmetric SOI is usually described by using an odd-parity function 
$\bm{\alpha}_{\bm{k}} = -  \bm{\alpha}_{-\bm{k}}$.   
In the two-valley model, however, 
they are described by a constant vector $\bm{\alpha}$ independent of the wave number $\bm{k}$ 
times the third Pauli matrix in valley space
$\hat{\rho}_z$.
SOI changes the sign under the spacial-inversion operator because $\hat{\rho}_x\hat{\rho}_z\hat{\rho}_x=-\hat{\rho}_z$.
The time-reversal operator is given by $\mathcal{T}=i\hat{\sigma}_{y}\hat{\rho}_{x}\mathcal{K}$, 
where $\mathcal{K}$ means taking the complex conjugation and the transformation of $\bm{k}=-\bm{k}$.
Particle-hole conjugation is represented by $\undertilde{\check{H}}_{\mathrm{N}} (\bm{k}) \equiv \check{H}^{*}_{\mathrm{N}} (-\bm{k})$ as usual.
We assume that two electrons at the different valleys form a spin-triplet Cooper pair.
The pair potential for such a pair is represented as
\begin{align}
  \check{\Delta}=\bm{d}\cdot\hat{\bm{\sigma}} i\hat{\sigma}_{y} i\hat{\rho}_{y},
\end{align}
where $\bm{d}$ can be a constant vector independent of $\bm{k}$.
The pair potential belongs to
spin-triplet, even-momentum-parity and odd-valley-parity symmetry class.
The pair potential also changes the sign under the spacial-inversion operator since $\hat{\rho}_xi\hat{\rho}_y\hat{\rho}_x=-i\hat{\rho}_y$.
It should be emphasized here that this does not refer to any specific materials.

By solving the Gor'kov equation for the Bogoliubov-de Gennes (BdG) Hamiltonian 
\begin{align}
    &[i\omega_n-H_{\mathrm{BdG}}]
    \begin{bmatrix}
      \check{\mathcal{G}} & \check{\mathcal{F}} \\
      - \undertilde{\check{\mathcal{F}}}  & - \undertilde{\check{\mathcal{G}}}
    \end{bmatrix}
    _{(\bm{k}, \omega_n)}
    =1, \label{eq:Gor'kov}\\
    &H_{\mathrm{BdG}}
    =
    \begin{bmatrix}
      \check{H}_{\mathrm{N}}(\bm{k}) & \check{\Delta} \\
      - \undertilde{\check{\Delta}}  & - \undertilde{\check{H}}_{\mathrm{N}}(\bm{k})
    \end{bmatrix} , \label{eq:BdG}
\end{align}
we obtain the Green's functions
\begin{align}
    \check{\mathcal{G}}(\bm{k}, \omega_n)&=
    \check{Z}^{-1}\bigl[(i\omega_n-\hat{\xi}_{\bm{k}})\hat{z}_N-(i\omega_n+\hat{\xi}_{\bm{k}})d^{2}\notag\\
    &\quad-(\hat{z}_N+d^{2})\bm{\alpha}\cdot\hat{\bm{\sigma}}\hat{\rho}_z
    -(\hat{z}_N-d^{2})\bm{V}\cdot\hat{\bm{\sigma}}\notag\\
    &\quad+2(\bm{\alpha}\cdot\bm{d})\bm{d}\cdot\hat{\bm{\sigma}}\hat{\rho}_z
    -2(\bm{V}\cdot\bm{d})\bm{d}\cdot\hat{\bm{\sigma}}\bigr], \label{eq:g}\\
	\check{\mathcal{F}}(\bm{k}, \omega_n)
    &= -\check{Z}^{-1}\bigl[
    (\omega_n^{2}+\hat{\xi}_{\bm{k}}^{2}+d^{2}-\alpha^{2}+V^{2})\bm{d}\cdot\hat{\bm{\sigma}} \notag \\
    &\quad +2(\bm{\alpha}\cdot\bm{d})\bm{\alpha}\cdot\hat{\bm{\sigma}}
    -2(\bm{V}\cdot\bm{d})\bm{V}\cdot\hat{\bm{\sigma}} \notag \\
    &\quad +2\hat{\xi}_{\bm{k}}\bm{\alpha}\cdot\bm{d}\hat{\sigma}_0\hat{\rho}_{z}
    +2i\hat{\xi}_{\bm{k}}\bm{V}\times\bm{d}\cdot\hat{\bm{\sigma}} \notag \\
    &\quad -2\omega_n\bm{\alpha}\times\bm{d}\cdot\hat{\bm{\sigma}}\hat{\rho}_{z}
    +2i\omega_n\bm{V}\cdot\bm{d}\hat{\sigma}_0 \notag\\
    &\quad-2i\bm{V}\times\bm{\alpha}\cdot\bm{d}\hat{\sigma}_0\hat{\rho}_{z}
    \bigr] i\hat{\sigma}_y i\hat{\rho}_y, \label{eq:f}
\end{align}
\begin{align}
	\check{Z} &= (\omega_n^{2}+\hat{\xi}_{\bm{k}}^{2}+d^{2}+\alpha^{2}-V^{2})^{2}\notag \\
    &\quad+4(\bm{V}\times\bm{\alpha})^{2}
    -4(i \omega_n\bm{V}+\hat{\xi}_{\bm{k}}\bm{\alpha}\hat{\rho}_z)^{2}\notag \\
    &\quad-4(|\bm{\alpha}\times\bm{d}|^{2}-|\bm{V}\times\bm{d}|^{2}), \label{eq:Z}\\
    \check{z}_N&=(i\omega_n+\hat{\xi}_{\bm{k}})^{2}-(\bm{V}-\bm{\alpha}\hat{\rho}_z)^{2}, \label{eq:zn}
\end{align}
where $\omega_n=(2n+1)\pi T$ is a Fermionic Matsubara frequency with $T$ being a temperature.
The magnetically active potentials induce Cooper pairs belong to various symmetry classes.
The first and second lines of Eq.~\eqref{eq:f} represent 
the paring correlation function belonging to
spin-triplet odd-valley-parity symmetry class and they are linked to the pair potential through
the gap equation,
\begin{align}
i \left(\bm{d} \cdot \hat{\bm{\sigma}} \, \hat{\sigma}_y \right)_{\alpha, \beta}
=& - T \sum_{\omega_n} \frac{1}{V_{\mathrm{vol}}} \sum_{\bm{k}}
\sum_{\gamma, \delta} \nonumber\\
& \times g_{\alpha\, \beta; \gamma\, \delta}\, 
\hat{\mathcal{F}}^{(y)}_{\gamma, \delta}(\bm{k}, \omega_n), \label{eq:gapeq}\\
g_{\alpha\, \beta; \gamma\, \delta}=& \sum_{\nu=x, y, z} g_\nu 
\left(i \hat{\sigma}_\nu \, \hat{\sigma}_y \right)_{\alpha, \beta}
\;
\left(i \hat{\sigma}_\nu \, \hat{\sigma}_y \right)^\ast_{\gamma, \delta}, \label{eq:g_conf}
\end{align}
where $\hat{\mathcal{F}}^{(y)}$ is the $\hat{\rho}_y$ component of the anomalous Green's function
and $\bm{g}=(g_x, g_y,g_z)$ represents the attractive interaction between two 
electrons at an $s$-wave channel.
All Cooper pairs in the two-valley model belong to even-momentum-parity $s$-wave symmetry class.
Hereafter, the pairing correlation forming the pair potential is referred to as principal 
pairing correlation. 
In contrast, the remaining correlations are induced by magnetically active potentials.
The two terms in the third line correspond to the spin-singlet even-valley-parity pairing correlation
and the spin-triplet odd-valley-parity pairing correlation, respectively. The former is induced by the SOI and 
the latter is induced by the Zeeman field.
The fourth line represents two odd-frequency pairing correlations: 
one is induced by the SOI and the other is induced by the Zeeman field.
The last term of Eq.~\eqref{eq:f} originates from the 
cooperative effect between $\bm{V}$ and $\bm{\alpha}$, and it belongs to spin-singlet 
even-valley parity class.

In this paper, we assume that $\bm{\alpha}$ and $\bm{d}$ are single-component vectors.
This simplification allows us to clarify the relationship between 
the stability of spin-triplet superconductivity
 and the relative vector configurations among $\bm{V}$, $\bm{\alpha}$ and $\bm{d}$.
In addition to this, it is possible to obtain the analytic expressions of the gap equation
and those of superfluid density because $\bm{\alpha}$ and $\bm{d}$ are independent of 
$\bm{k}$. 
These analytical results enable us to understand how induced subdominant 
pairing correlations affect the stability of the superconducting states, which is 
the advantage of the two-valley model.
Furthermore, we restrict the relative orientations of the three vectors 
to be either perpendicular or parallel to one another.
The resulting five configurations are summarized in Table.~\ref{tab:stability}.
\begin{table*}[t]
  \centering
  \caption{The relation between the stability of superconductivity 
  and the relative alignment among the three vectors.
  The last three terms in Eq.~\eqref{eq:f} correspond to the induced pairing correlations.
  The column ``odd-$\omega$'' (``even-$\omega$'' ) shows the induced pairing correlations
  belonging to odd-frequency (even-frequency) symmetry class.
  The symbols ``$\times$'' and ``$\circ$'' indicate that the corresponding potential 
  destabilizes and stabilizes the superconducting state, respectively.
  The symbol ``-'' means no effect. 
  In case 5, ``$\triangle$'' means that SOI destabilize (stabilize ) the superconducting 
  state for small $\bm{V}$ (large $\bm{V}$).}
  \label{tab:stability}
  \setlength{\tabcolsep}{10pt}
  \renewcommand{\arraystretch}{1.3}
  \begin{tabular}{cccccccc}
    \toprule
    Case & $\bm{V}$-$\bm{d}$ & $\bm{\alpha}$-$\bm{d}$ & $\bm{V}$-$\bm{\alpha}$ &odd-$\omega$& even-$\omega$ 
	& Effect of $\bm{\alpha}$ & Effect of $\bm{V}$\\
    \hline
    1 & $\perp$ & $\parallel$ & $\perp$ & & & - & -\\
    2 & $\parallel$ & $\parallel$ & $\parallel$ &$\bm{V}\cdot\bm{d}\,  \hat{\rho}_y$ &  & - & $\times$ \\
    3 & $\perp$ & $\perp$ & $\parallel$ & $\bm{\alpha}\times \bm{d} \cdot \hat{\bm{\sigma}}\, \hat{\rho}_x$ & & $\times$ & - \\
    4 & $\parallel$ & $\perp$ & $\perp$ &
 $\bm{\alpha}\times \bm{d} \cdot \hat{\bm{\sigma}}\, \hat{\rho}_x$, 
 $\bm{V}\cdot\bm{d}\,  \hat{\rho}_y$& & $\times$ & $\times$ \\
    5 & $\perp$ & $\perp$ & $\perp$ & $\bm{\alpha}\times \bm{d} \cdot \hat{\bm{\sigma}}\, \hat{\rho}_x$& 
	$\bm{\alpha}\times \bm{V} \cdot \bm{d}\, \hat{\rho}_x$& $\triangle$ & $\circ$ \\
    \toprule
  \end{tabular}
\end{table*}

The transition temperature $T_c$ to the superconducting phases 
is a tool that measures the stability or instability of the 
superconducting state. 
Namely, the higher $T_c$ of a superconducting state, the more stable it is.
The transition temperatures are calculated in two ways:
by solving the gap equation in Eq.~\eqref{eq:gapeq} and by finding
the minimum of the free-energy,
\begin{align}
  \Omega_S&=-\frac{1}{2}T\frac{1}{V_{\mathrm{vol}}}\sum_{\bm{k}}\sum_{\sigma,\rho=\pm1}
  \ln \left[2\cosh \left(\frac{E_{\bm{k},\sigma,\rho}}{2T}\right)\right]+\frac{d^{2}}{g}, \label{eq:free_en}\\
 & \approx  a \, d^2 + b \, d^4 + \cdots, \label{eq:gl}
\end{align}
where $E_{\bm{k},\sigma,\rho}$ is the eigen energy of the BdG Hamiltonian.
Two transition temperatures are identical to each other when the transition to the superconducting state is a second-order.
For a first-order transition, the phase boundary is obtained from the free-energy minima.
To understand the mechanisms that stabilize/destabilize the superconducting states,
 we discuss the superfluid weight $Q$ which enters
the electric current in the presence of the vector potential $\bm{A}$ as
\begin{align}
\bm{j} = - \frac{n e^2 Q}{m c} \bm{A}, \label{eq:london}
\end{align}
 where $n$ is the electron density. 
The superfluid weight is defined as~\cite{agd}
\begin{align}
  Q&=Q_{\mathcal{G}}+Q_{\mathcal{F}}, \\
  Q_{\mathcal{G}}&=\frac{1}{2}\, T\sum_{\omega_n}\int d\xi\,
  \mathrm{Tr}[\check{\mathcal{G}}\check{\mathcal{G}}-\check{\mathcal{G}}^{\mathrm{N}}\check{\mathcal{G}}^{\mathrm{N}}]_{(\bm{k}, \omega_n)},\\
  Q_{\mathcal{F}}&=\frac{1}{2}\, T\sum_{\omega_n}\int d\xi\,
  \mathrm{Tr}[-\undertilde{\check{\mathcal{F}}}\check{\mathcal{F}}]_{(\bm{k}, \omega_n)}, \label{eq:qf_def} 
\end{align}
where $\check{\mathcal{G}^{\mathrm{N}}}$ is the Green's function in the normal state.
Since the relation $Q_{\mathcal{G}} = Q_{\mathcal{F}}$ 
as shown in Appendix~\ref{secap:qg_qf}, we focus only on $Q_{\mathcal{F}}$.
In Appendix~\ref{secap:qf}, we show the general expressions of $Q_{\mathcal{F}}$ which is decomposed 
into the contributions from each pairing correlation in Eq.~\eqref{eq:f}.
Together with the Maxwell's equation, Eq.~\eqref{eq:london} explanes the Meissner effect. 
The superfluid weight characterizes the penetration length of a magnetic field as 
$\lambda \propto {Q_{\mathcal{F}}}^{-1/2}$.  
The superconducting states with larger $Q_{\mathcal{F}}$ are more stable.
Finally, we discuss the relation between $Q_{\mathcal{F}}$ just below $T_c$ and a 
coefficient of 
the Ginzburg-Landau free-energy~\eqref{eq:gl}.
It is well established that the coefficient $a$ is proportional to
the linearized gap equation. 
Since the superfluid weight is proportional to $d^2$, the dependence of $Q_{\mathcal{F}}$ on temperatures 
is strongly governed by the dependence of $d$ on temperatures.
Here, we evaluate the superfluid weight
\begin{align}
  q (T,V_c)
  =
  \frac{Q_{\mathcal{F}}[V_{c} (T), T]}{d^{2}} \bigg|_{d \to 0}  \label{eq:sqdef}
\end{align}
instead of $Q_{\mathcal{F}}$ itself, where $V_c= \mu_{\mathrm{B}} H_c$ corresponds 
to the critical Zeeman potential on the phase boundary. 
The coefficient $b$ is proportional to the superfluid weight $q(T_c, V_c)$~\cite{sato:prb2024}.
Therefore, the transition becomes a first-order for $Q(T_c, V_c) <0$.

\section{Results} \label{sec:results}

Before turning into superconducting states in each configuration, we briefly summarize 
the outline of this section.
There are four types correlations in the anomalous Greens' function.
The first line in Eq.~\eqref{eq:f} is the principal component that forms the pair potential 
through the gap equation.
The two correlations in the second line in Eq.~\eqref{eq:f} belong to 
the same symmetry class as the principal component and contribute to the pair 
potential. Therefore, we recognize them as parts of the principal component.  
In Case 1, we discuss the roles of the two correlations at the third line proportional to $\hat{\xi}_{\bm{k}}$. 
They do not affect the stability of superconducting states.
The remaining three components are the induced correlations that 
stabilize or destabilize the superconducting states.
They belong to different symmetry class from that in the principal component.
As shown in Table~\ref{tab:stability},  a Zeeman field, SOI, and both of them induce the odd-frequency 
pairing correlations in Cases 2, 3, and 4, respectively. 
Odd-frequency Cooper pairs make the superconducting state unstable because they decrease the 
superfluid weight.  As a result, $T_c$ decreases with increasing the amplitude of odd-frequency pairing 
correlations. The $H$-$T$ phase diagrams for Cases 2 - 4 are shown in Fig.~\ref{fig:case2_4}.
The configuration in Case 5 causes ZFIS as shown in  the phase diagram in Fig.~\ref{fig:case5}.
The spin-singlet pairing correlation at the last term in Eq.~\eqref{eq:f} stabilizes the spin-triplet 
superconductivity in high Zeeman fields.

\subsection{Case 1} \label{sub1}
We begin our discussion with Cace 1 in Table~\ref{tab:stability}.
It has been known that either the SOI $\bm{\alpha} \parallel \bm{d}$ or the Zeeman field $\bm{V} \perp \bm{d}$ 
does not affect the stability of superconducting states.
Such situation is described by the absence of the induced pairing correlations
at the last three terms in the anomalous Green's function in Eq.~\eqref{eq:f}.
The first term at the third line in Eq.~\eqref{eq:f} represents 
the spin-singlet even-valley-parity pariring correlation and all the remaining 
components belong to spin-triplet odd-valley-parity symmetry class.
Here we mainly explain the roles of the pairing correlations proportional to 
$\hat{\xi}_{\bm{k}}$. 
The anomalous Green's function and the eigen energy are obtained as
\begin{align}
  \check{\mathcal{F}}_1&(\bm{k}, \omega_n)
  = -\check{Z}^{-1}_1\bigl[
  (\omega_n^{2}+\hat{\xi}_{\bm{k}}^{2}+d^{2}+\alpha^{2}+V^{2})\bm{d}\cdot\hat{\bm{\sigma}}\notag\\
  &
  +2i\hat{\xi}_{\bm{k}}\bm{V}
  \times\bm{d}\cdot\hat{\bm{\sigma}}
  +2\hat{\xi}_{\bm{k}}\bm{\alpha}
  \cdot\bm{d}\hat{\sigma}_0\hat{\rho}_z
  \bigr] i\hat{\sigma}_y i\hat{\rho}_y, \label{eq:f_1} \\
  \check{Z}_1 
  &=\left[\left(\hat{\xi}_{\bm{k}}+\sqrt{V^{2}+\alpha^{2}}\right)^{2}+\omega_n^{2}+d^{2}\right]\notag\\
  &\quad\quad\quad\times\left[\left(\hat{\xi}_{\bm{k}}-\sqrt{V^{2}+\alpha^{2}}\right)^{2}+\omega_n^{2}+d^{2}\right], \label{eq:Z_1}\\
  E_{\bm{k},\sigma,\rho}&=
  \sqrt{(\xi_{\rho}+\sigma\sqrt{V^{2}+\alpha^{2}})^{2}+d^{2}}. \label{eq:eig_1}
\end{align}
Here $\xi_{\rho} =(\bm{k} - \rho \bm{K})^{2}/2m -\mu $ with $\rho = \pm 1$ are dispersion 
at the two valleys and $\sigma = \pm 1$ is derived from the spin degree of freedom.
The first term in Eq.~\eqref{eq:f_1} is the principal correlation. 
The summation over wavenumber is necessary to
reach the gap equation, 
\begin{align}
\frac{1}{V_{\mathrm{vol}}} \sum_{\bm{k}} \check{\mathcal{F}}_1(\bm{k}, \omega_n) 
=& - N_0 \int d \xi \frac{  \bm{d} \cdot \hat{\bm{\sigma}} \hat{\sigma}_y }{ \xi^2 + \omega_n^2 + d^2}, \\
=& - \frac{\pi N_0}{\sqrt{ \omega_n^2 + d^2}} \bm{d} \cdot \hat{\bm{\sigma}} \, \hat{\sigma}_y,
\end{align}
where $N_0$ is the density of states in the normal state at the Fermi level per spin.
The shift of $\bm{k}$ to $\bm{k} \pm \bm{K}$ for $\xi_{\pm}$
does not change the results because $\bm{d}$ is independent of $\bm{k}$. 
The change of the integration variable $\xi$ to $\xi \pm  \sqrt{V^{2}+\alpha^{2}} $ 
leads to the last line. 
In Case 1, both a Zeeman field and a SOI modify 
the normal state dispersion only slightly. 
Such effects are eliminated by changing the integration variable. 
The resulting gap equation 
\begin{align}
  1
  &=2\pi gN_0T\sum_{\omega_n>0}\frac{1}{\sqrt{\omega_n^{2}+d^{2}}}, \label{eq:gap_1}
\end{align}
is equivalent to that for a conventional SC.
Neither a Zeeman field nor a SOI changes $T_c$ at all.

All the three terms in Eq.~\eqref{eq:f_1} contibute to the superfluid weight.
The results
\begin{align}
Q_{\mathcal{F},1} = {\pi} T\sum_{\omega_n} \frac{{d}^2}{ (\omega_n^2+d^2)^{3/2} }
 =  Q_{\mathrm{BCS}},
\end{align}
indicate that the superfluid weight at $\bm{\alpha}=\bm{H}=0$ is recovered. 
Thus the pairing correlations proportional to $\hat{\xi}$ do not affect the stability of the
superconducting state at all.
This conclusion is also true in all the configurations.
Therefore, these correlations are not listed in Table~\ref{tab:stability}.

\subsection{Case 2}\label{sub2}

\begin{figure}[t]
  \centering
  \includegraphics[width=\linewidth]{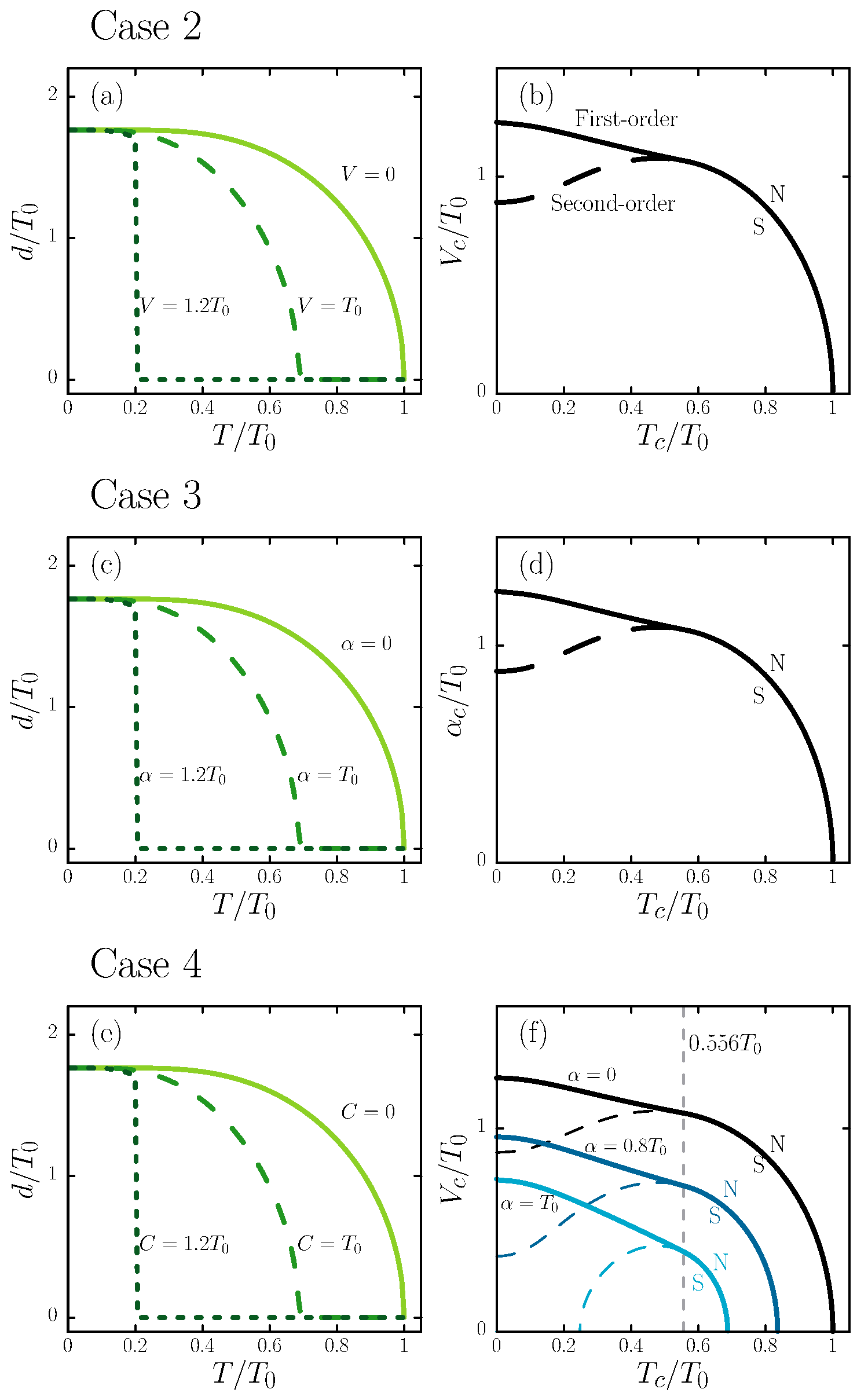}
  \caption{ The results for case 2, those for Case 3 and those for Case 4 
  are shown in (a)- (b),  (c)-(d) and (e)-(f), respectively.
The amplitude of the pair potential is plotted as a function of temperatures in (a), (c) and (e).
 The phase diagram is shown in (b), (d), and (f), where
  the solid and dashed lines are obtained from the minima of Eq.~\eqref{eq:free_en} and 
 the solution of Eq.~\eqref{eq:linearized_2}, respectively. 
  The results for Case 2 in (a) and (b) are independent of $\alpha$.
 The results for Case 3 are identical to those for Case 2 with interchanging 
  the two magnetic potentials $ V \leftrightarrow  \alpha$. 
 The results for Case 4 in (e) are identical to those for Case 2 in (a) by 
 replacing  $ V$ by  $C=\sqrt{V^2+  \alpha^2}$. 
 In (f), a vertical broken line indicates $T_c=0.556T_0$ below which the transition becomes 
 a first-order because induced odd-frequency pairs decrease the superfluid weight to be negative.
This property is common in (b), (d) and (f). 
  }
  \label{fig:case2_4}
\end{figure}

The anomalous Green's function in case 2 with $\bm{d} \parallel \bm{\alpha} \parallel \bm{V}$ 
is obtained as
\begin{align}
  \check{\mathcal{F}}_2(&\bm{k}, \omega_n)
  = -\check{Z}^{-1}_2\bigl[
  (\omega_n^{2}+\hat{\xi}_{\bm{k}}^{2}+d^{2}+\alpha^{2}-V^{2})\bm{d}\cdot\hat{\bm{\sigma}}\notag\\
  &\quad\quad+2i\omega_n\bm{V}\cdot\bm{d}\hat{\sigma}_0
  +2\hat{\xi}_{\bm{k}}\bm{\alpha}
  \cdot\bm{d}\hat{\sigma}_0\hat{\rho}_z
  \bigr] i\hat{\sigma}_y i\hat{\rho}_y, \label{eq:f_2} \\
  \check{Z}_2 
  &=\left[(\hat{\xi}_{\bm{k}}+\alpha\hat{\rho}_z)^{2}+(\omega_n+iV)^{2}+d^{2}\right]\notag\\
  &\quad\quad\quad\times\left[(\hat{\xi}_{\bm{k}}-\alpha\hat{\rho}_z)^{2}+(\omega_n-iV)^{2}+d^{2}\right]. \label{eq:Z_2}
\end{align}
The free-energy is obtained by using the eigen energies of the BdG Hamiltonian
\begin{align}
  E_{\bm{k},\sigma,\rho}&=
  \sqrt{(\xi_{\rho}+\sigma\rho\alpha)^{2}+d^{2}}+\sigma V. \label{eq:eig_2}
\end{align}
The first term of Eq.~\eqref{eq:f_2} represents the principal pairing correlation.
The last term does not affect the stability as discussed in Case 1.
The odd-frequency pairing correlation induced by the Zeeman field at the second term 
makes the superconducting states unstable. 
The gap equation is derived as
\begin{align}
 W_d(T,V)=&1,\label{eq:gap_2}
\end{align}
with
\begin{align} 
 W_d(T,X)=&2\pi gN_0T\sum_{\omega_n>0}
  \mathrm{Re}\left[\frac{1}{\sqrt{(\omega_n+iX)^{2}+d^{2}}}\right]. \label{eq:w_def}
\end{align}
Since the gap equation share the common structure here and 
the Cases 3 and 4, we represent it in a unified way. 
Equation~\eqref{eq:gap_2} is identical to the gap equation for a spin-singlet SC
 in the presence of a Zeeman field.
The energy eigenvalues in Eq.~\eqref{eq:eig_2} also indicate the effects 
of the two magnetically active potentials on superconducting states.
The SOI $\alpha$ only renormalizes the chemical potential, whereas 
a Zeeman field $V$ reduces the gap for $\sigma=-1$.
 In Fig.~\ref{fig:case2_4}(a), the dependence of the amplitude of the pair potential 
 on temperature is shown for several choices of Zeeman potentials $V$, 
 $T_0$ is the transition temperature at $V=\alpha=0$.
The Zeeman field suppresses superconductivity, whereas the SOI does not change $T_c$ at all.
The results for $V=1.2T_0$ shows the jump of $d$ at the transition temperature.
Namely, the transition becomes a first-order~\cite{maki_tsuneto:ptp1964,sarma:jpcs1963}. 
Although $T_c$ decreases with increasing Zeeman fields, the amplitude of the 
pair potential at $T=0$ remain unchanged~\cite{sarma:jpcs1963}.
The solution of $d$ is obtained from two methods: searching minimum of the free-energy 
in Eq.~\eqref{eq:free_en}
and solving the gap equation in Eq.~\eqref{eq:gap_2}.
The two solutions are identical to each other for a second-order transition, whereas 
they deviate from each other for a first-order transition.

The linearized gap equation results in 
\begin{align}
U_0(T_c, V) =& 0, \label{eq:linearized_2}
\end{align}
with
\begin{align}
U_0(T,X)=&
 \ln \frac{T}{T_0}+2\pi T\sum_{\omega_n>0}\frac{1}{\omega_n}
 \left[1-\frac{\omega_n^{2}}{\omega_n^{2}+X^{2}}\right]. \label{eq:u_def}
\end{align} 
This result agrees with that obtained in Ref.~\cite{Mockli:prb2020}.
The second term in Eq.~\eqref{eq:u_def} increases from 0 with increasing $X$, which leads to 
the suppression of $T_c$ from $T_0$. 
Thus, Eqs.~\eqref{eq:linearized_2} and \eqref{eq:u_def} indicate that 
$T_c$ decreases with increasing Zeeman fields $V$.
The $H$-$T$ phase diagram in Fig.~\eqref{fig:case2_4}(b) is obtained in two ways: 
solving the linearized gap equation as shown with a broken line and finding the minimum of the free-energy as shown 
with a solid line.
These two lines deviate from each other at low temperature for $T<0.556T_0$~\cite{maki_tsuneto:ptp1964,sarma:jpcs1963}. 
There are two contributions to the superfluid weight,
\begin{align}
  q(T, X) &= q_{0}(T, X) + q_{\mathrm{odd}}(T, X) , \label{eq:samll_q2}\\
  q_{0}(T, X)
  &=
  2 \pi T \sum_{\omega_{n} > 0}
  \frac{2 \omega^{4}_{n} - \omega^{2}_{n}  X^{2} +  X^{4}}
  {2 \omega_{n} (\omega^{2}_{n} + X^{2})^{3}} , \label{eq:samll_q2e}\\
  q_{ \mathrm{odd}}(T, X)
  &=
  - 2 \pi T \sum_{\omega_{n} > 0}
  \frac{5 \omega^{2}_{n}  X^{2} +  X^{4} }
  {2 \omega_{n} (\omega^{2}_{n} + X^{2})^{3}}. \label{eq:samll_q2o}
\end{align}
As shown in Appendix~\ref{secap:qf},
$q_0$ is the superfluid weight due to the principal correlations.
The weight due to odd-frequency correlations $q_{ \mathrm{odd}}$ is
 derived from Eq.~\eqref{eq:Q3}.   
As already confirmed in the previous paper~\cite{sato:prb2024},
 the superfluid weight
 $q(T, V_c)$ \textsl{along the transition line} in Fig.~\ref{fig:case2_4}(b) 
 becomes negative for $T<0.556T_0$. 
 As a result, the transition to the superconducting state becomes a first-order 
 as indicated by the results with $V=1.2T_0$ in Fig.~\ref{fig:case2_4}(a). 
Odd-frequency Cooper pairs induced by a Zeeman field make 
the superconducting state unstable as listed in Table~\ref{tab:stability}. 
The zero superfluid weight $q=0$ also indicates the critical point for a nonuniform 
superconducting state characterized by a finite center-of-mass momentum of a Cooper pair~\cite{sato:fflo2025}.
Such a finite-momentum state in a conventional SC under Zeeman field is known as  
Fulde-Ferrell-Larkin-Ovchinnikov (FFLO) state~\cite{fulde_fflo,larkin_fflo}.
Thus finite-momentum superconducting states are possible also for spin-triplet SCs.
This conclusion can be applied also to Cases 3 and 4, where the 
SOI $\bm{\alpha} \perp \bm{d}$ generates the odd-frequency pairing correlations.

\subsection{Case 3}\label{sub3}

The anomalous Green's function and the eigen energy for case 3 with $\bm{\alpha}\perp \bm{d}$, 
$\bm{V} \perp \bm{d}$, and $\bm{\alpha} \parallel \bm{V}$ become
\begin{align}
  \check{\mathcal{F}}_3(&\bm{k}, \omega_n)
  = -\check{Z}^{-1}_3\bigl[
  (\omega_n^{2}+\hat{\xi}_{\bm{k}}^{2}+d^{2}-\alpha^{2}+V^{2})\bm{d}\cdot\hat{\bm{\sigma}}\notag\\
  &\quad\quad-2\omega_n\bm{\alpha}\times\bm{d}\cdot\hat{\bm{\sigma}}\hat{\rho}_z
  +2i\hat{\xi}_{\bm{k}}\bm{V}
  \times\bm{d}\cdot\hat{\bm{\sigma}}
  \bigr] i\hat{\sigma}_y i\hat{\rho}_y, \label{eq:f_3} \\
  \check{Z}_3 
  &=\left[(\hat{\xi}_{\bm{k}}+V)^{2}+(\omega_n+i\alpha\hat{\rho}_z)^{2}+d^{2}\right]\notag\\
  &\quad\quad\quad\times\left[(\hat{\xi}_{\bm{k}}-V)^{2}+(\omega_n-i\alpha\hat{\rho}_z)^{2}+d^{2}\right], \label{eq:Z_3}\\
  E_{\bm{k},\sigma,\rho}&=
  \sqrt{(\xi_{\rho}+\sigma V)^{2}+d^{2}}+\sigma\rho\alpha. \label{eq:eig_3}
\end{align}
In contrast to case 2, the odd-frequency correlation is induced by the SOI instead of a Zeeman field
 as shown in the second term. 
The roles of the two potentials in Case 3 are the reverse of those in Case 2.
Therefore, we explain the results for Case 3 with paying attention to the difference from those in Case 2.
The gap equation and it in the linear regime
are $ W_d(T,\alpha)=1$ and $U_0(T_c, \alpha)=0$, respectively.
The temperature dependence of the pair potential in Case 3 is shown in 
Fig.~\ref{fig:case2_4}(c) for several choices of $\alpha$. 
The phase diagrams in Case 3 are shown in Fig.~\ref{fig:case2_4}(d).
The vertical axis is the critical value of the SOI.
These curves are exactly the same as those in Fig.~\ref{fig:case2_4}(b). 
The superfluid weight is also described by Eqs.~\eqref{eq:samll_q2}-\eqref{eq:samll_q2o} with $X=\alpha$, 
where $q_{\mathrm{odd}}$ is derived from $q_{\bm{\alpha}\perp \bm{d}}$ in Eq.~\eqref{eq:Q4} and 
$\alpha_c$ is the critical value of SOI on the phase boundary. 
The instability of the superconducting state is characterized by the appearance 
of odd-frequency pairs as listed in Table~\ref{tab:stability}. 

The odd-frequency pairing correlations induced by the SOI 
$\bm{\alpha} \perp \bm{d}$ decrease the superfluid weight to be negative. 
This fact indicates the possibility of a finite-momentum spin-triplet superconducting state 
under the SOI.

\subsection{Case 4} \label{sub4}

The anomalous Green's function in Case 4 and the eigen energy are given by
\begin{align}
  \check{\mathcal{F}}_4(&\bm{k}, \omega_n)
  = -\check{Z}^{-1}_4\bigl[
  (\omega_n^{2}+\hat{\xi}_{\bm{k}}^{2}+d^{2}-\alpha^{2}-V^{2})\bm{d}\cdot\hat{\bm{\sigma}}\notag\\
  &\quad\quad-2\omega_n\bm{\alpha}\times\bm{d}\cdot\hat{\bm{\sigma}}\hat{\rho}_z
  +2i\omega_n\bm{V}
  \cdot\bm{d}\hat{\sigma}_0
  \bigr] i\hat{\sigma}_y i\hat{\rho}_y, \label{eq:f_4} \\
  \check{Z}_4 
  &=\left[\hat{\xi}_{\bm{k}}^{2}+\left(\omega_n+i C\right)^{2}+d^{2}\right]\notag\\
  &\quad\quad\quad\times\left[\hat{\xi}_{\bm{k}}^{2}+\left(\omega_n-i C\right)^{2}+d^{2}\right], \label{eq:Z_4}
  \end{align}
  \begin{align}
  E_{\bm{k},\sigma,\rho}&=
  \sqrt{\xi_{\rho}^{2}+d^{2}}+\sigma C, \label{eq:eig_4}\\
  C&= \sqrt{\alpha^2+V^2},
\end{align}
where $C$ represents the amplitude of the combined potential.
The two odd-frequency pairing correlations appear in Case 4 as listed in Table~\ref{tab:stability}.
One is derived from the SOI and the other is induced by the Zeeman field.
They are expected to destabilize the superconducting state.
The eigenvalue in Eq.~\eqref{eq:eig_4} also show that both $\alpha$ and $V$ reduce the gap for $\sigma=-1$.
The gap equation and that in the linear regime become
$W_d(T, C ) = 1$ and $U_0(T_c, C )=0$, respectively.
The pair potential is plotted as a function of temperatures for several $C$ in Fig.~\ref{fig:case2_4}(e).
The $H$-$T$ phase diagram at several values of $\alpha$ is shown in Fig.~\ref{fig:case2_4}(f).
The critical magnetic field decreases with increasing $\alpha$.
The discrepancy between the first-order transition points shown with solid lines and 
the second-order transition points shown with broken lines becomes more remarkable for 
larger $\alpha$. 
The superfluid weight is represented by Eq.~\eqref{eq:samll_q2} with $X=C_c$, where 
$C_c$ is the critical value of
the combined potential on the phase boundary in Fig.~\ref{fig:case2_4}(f).
The two odd-frequency pairing correlations contribute to the superfluid weight
of odd-frequency pairs
\begin{align}
q_{\mathrm{odd}} = q_{\bm{V}\parallel \bm{d}} + q_{\bm{\alpha}\perp \bm{d}}.
\end{align} 
Although the critical value of a Zeeman field in Fig.~\ref{fig:case2_4}(f) depends on 
the amplitude of the SOI, the transition always becomes 
a first-order at $T \approx 0.556T_0$ for the all results.

\subsection{Case 5}\label{sub5}

\begin{figure}[t]
  \centering
  \includegraphics[width=\linewidth]{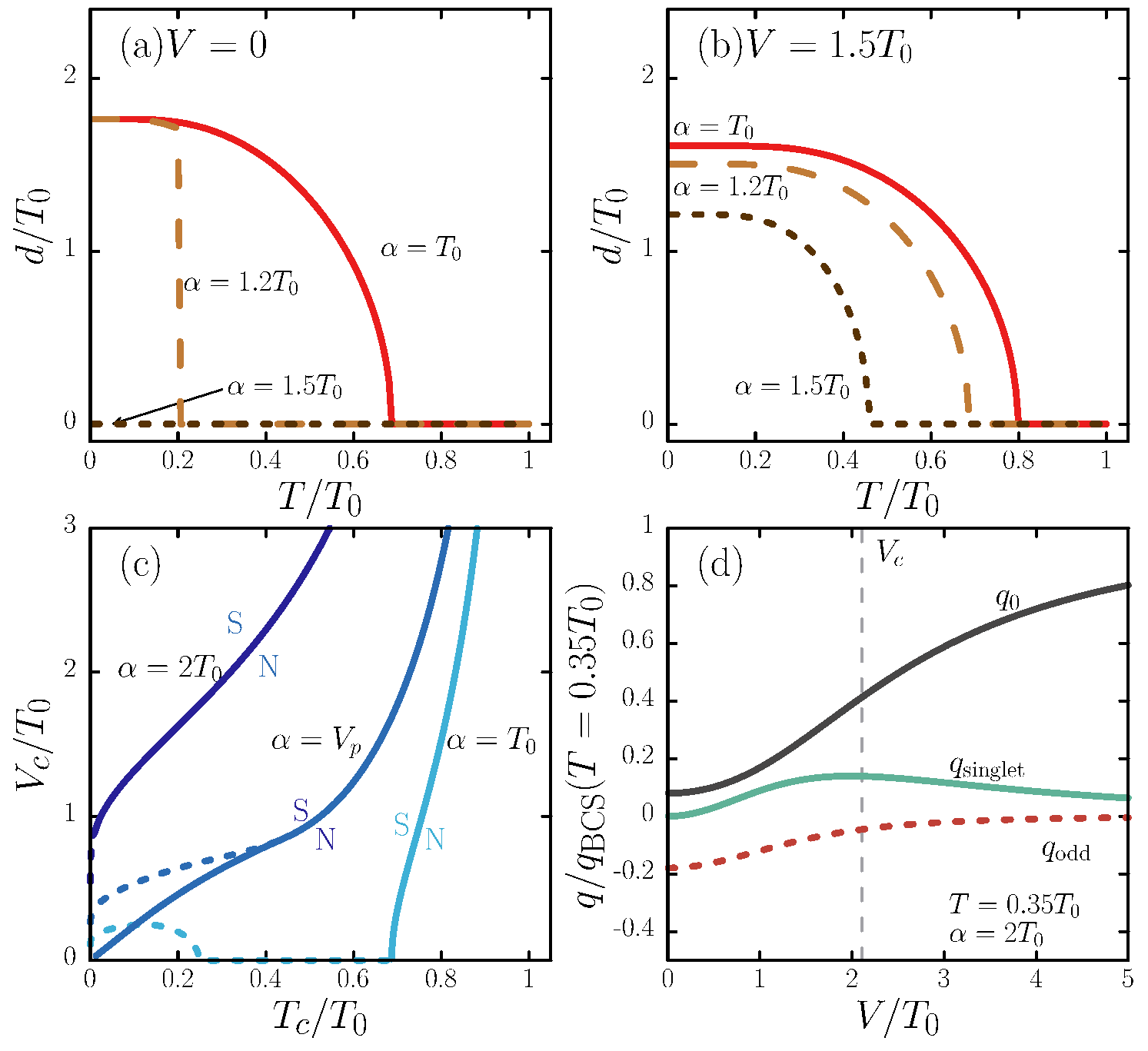}
  \caption{
  The results for Case 5.
  The amplitudes of the pair potential are plotted as a function of temperature 
 for several $\alpha$ at $V=0$ in (a) and at $V=1.5T_0$ in (b).
  The phase diagrams are shown for $\alpha=T_0$, $V_p$ and $2T_0$.
  The solid and dashed lines indicate the phase boundary of the first-order transition 
  and that of the second-order transition, respectively.
  The superfluid weights are plotted as a function of $V$ at $\alpha=2T_0$ 
  and $T=0.35T_0$, where a vertical broken line indicate the critical Zeeman field.
  }
  \label{fig:case5}
\end{figure}

In Cases 1-4, the instability of superconducting states is well described by the 
amplitude of the magnetic potentials that generate odd-frequency Cooper pairs.
As listed on Table.~\ref{tab:stability}, such amplitudes are 0, $V$, $\alpha$, and 
$\sqrt{\alpha^2+V^2}$, in Subsections~\ref{sub1}, \ref{sub2}, \ref{sub3}, and \ref{sub4}, respectively.
Roughly speaking, the effects of a SOI and those of a Zeeman field are additive in these cases.
In Case 5, the effects of these two potentials on the spin-triplet superconducting state are
qualitatively different from those in the previous cases.
Although the SOI suppresses superconductivity at zero Zeeman field,  
it enhances superconductivity in high Zeeman fields. 
Consequently, we will show that the superconductivity is induced by Zeeman fields.

The anomalous Green's function and the eigenvalue are given by
\begin{align}
  \check{\mathcal{F}}_5(&\bm{k}, \omega_n)
  = -\check{Z}_5^{-1}\bigl[
  (\omega_n^{2}+\hat{\xi}_{\bm{k}}^{2}+d^{2}-\alpha^{2}+V^{2})\bm{d}\cdot\hat{\bm{\sigma}}\notag\\
  &\quad\quad-2\omega_n\bm{\alpha}\times\bm{d}\cdot\hat{\bm{\sigma}}\hat{\rho}_z
  +2i\hat{\xi}_{\bm{k}}\bm{V}
  \times\bm{d}\cdot\hat{\bm{\sigma}}\notag\\
  &\quad\quad-2i\bm{V}\times\bm{\alpha}\cdot\bm{d}\hat{\sigma}_0\hat{\rho}_z
  \bigr] i\hat{\sigma}_y i\hat{\rho}_y, \label{eq:f_5} \\
  \check{Z}_5 
  &=(\omega_n^{2}+\hat{\xi}_{\bm{k}}^{2}+d^{2}-\alpha^{2}+V^{2})^{2}\notag\\
  &\quad\quad\quad+4V^{2}\alpha^{2}+4\omega_n^{2}\alpha^{2}-4\hat{\xi}_{\bm{k}}^{2}V^{2}, \label{eq:Z_5}\\
  E_{\bm{k},\sigma,\rho}&=
  \sqrt{\xi_{\rho}^{2}+d^{2}+C^{2}+2\sigma\sqrt{ C^{2}\, \xi_{\rho}^{2}+d^{2}\alpha^{2}}}. \label{eq:eig_5}
\end{align}
The first term in Eq.~\eqref{eq:f_5} is the principal correlation function that forms the pair potential.
The second term represents odd-frequency pairing correlation induced by the 
SOI as listed on Table~\ref{tab:stability}. 
As already discussed in case 1, the pairing correlation of the third term 
contribute to the superfluid weight $Q_0$.
Finally, the last term is the correlation function of spin-singlet even-valley-parity Cooper pairs 
induced by the interplay between the SOI and the Zeeman field. 
Since it belongs to even-frequency symmetry class, 
such spin-singlet Cooper pairs stabilize the superconducting state even in high Zeeman fields.
The energy eigenvalue in Eq.~\eqref{eq:eig_5} also suggests that the roles of the Zeeman field and 
those of the SOI are not separable 
into simple physical pictures such as a chemical potential shift or gap closing.
Although the gap equation
\begin{widetext}
\begin{align}
  1&=2\pi gN_0T
  \sum_{\omega_n>0}
  \mathrm{Re}\left[\frac{1+iV^{2}/\sqrt{\omega_n^{2}\alpha^{2}+(\omega_n^{2}+d^{2})V^{2}}}{\sqrt{\omega_n^{2}+d^{2}-\alpha^{2}-V^{2}+2i\sqrt{\omega_n^{2}\alpha^{2}+(\omega_n^{2}+d^{2})V^{2}}}}\right]. \label{eq:gap_5}
\end{align}
\end{widetext}
has a complicated structure due to the cross term between $\alpha$ and $V$,  
the linearized gap equation has a simpler expression as
\begin{align}
  \ln \frac{T}{T_0}&+2\pi T\sum_{\omega_n>0}\frac{1}{\omega_n}\left[1-\frac{\omega_n^{2}+V^{2}}{\omega_n^{2}+\alpha^{2}+V^{2}}\right]\notag\\
  &=0. \label{eq:linearized_5}
\end{align}
  The results enable us to understand the roles of a Zeeman field.
At $V=0$, the large SOI $\alpha > \mu_{\mathrm{B}} H_{\mathrm{P}}$ deletes superconductivity according to 
the argument in Case 3.
The Zeeman field screens such negative effects due to the SOI and removes them for $V \gg \alpha$.  
The pair potential  are plotted as a function of temperatures
for several $\alpha$ in Figs.~\ref{fig:case5}(a) for $V=0$ and (b) for $V=1.5T_0$. 
The comparison between  (a) and (b) makes clear the effects of Zeeman fields on the superconducting state. 
At $V=0$ in (a), the results for $\alpha=T_0$ show a second-order transition at $T=0.67T_0$ and 
those for $\alpha=1.2T_0$ indicate a first-order transition at $T=0.2T_0$. 
Superconductivity is absent for $\alpha=1.5T_0$ because the amplitude of the SOI goes over the Pauli limit.
 The results in (b) show that the Zeeman field stabilizes the superconducting state 
 and increase $T_c$ for all $\alpha$.
The first-order transition at $\alpha=1.2T_0$ in (a) is changed to a second-order in (b).
Finally, the superconducting phase appears only in the Zeeman field for $\alpha=1.5T_0$.

The $H$-$T$ phase diagram is shown in Fig.~\ref{fig:case5}(c) for several choices of $\alpha$.
For $\alpha=T_0$,  $T_c$  increases almost linearly with increasing Zeeman fields from $0.67 T_0$ at $V=0$. 
The phase boundary due to the second-order transition shown with a broken line indicates 
the transition to the normal state at very low temperature. 
However, the appearance of the low-temperature normal phase is unphysical because
the entropy of such normal phase must be smaller that that of the superconducting phase.
The phase boundary drawn by the free-energy minimum should be realized.
At the Pauli limit of $\alpha = V_{\mathrm{P}}$, $V_c$ increases almost linearly with $T_c$ for $T_c < 0.5T_0$, 
where the transition to the superconducting state is a first-order.
For $\alpha=2 T_0$, the superconductivity is absent at $V=0$ because of $\alpha>V_{\mathrm{P}}$.
The superconducting phase appears around $T \approx 0$ for $V>0.6T_0$ and extends into the high-temperature region
 with increasing $V$. 
 The results indicate ZFIS. 
 The transition to the superconducting phase is a second-order.

The analysis of superfluid weight 
\begin{align}
q= q_0 + q_{\mathrm{odd}} + q_{\mathrm{singlet}}
\end{align}
is helpful for us to understand the mechanism of the Zeeman-field-induced superconducting state.
The expression of these terms are supplied in Appendix~\ref{secap:q5},
where $q_{\mathrm{odd}} $ and $q_{\mathrm{singlet}}$ originate from 
Eqs.~\eqref{eq:Q4} and \eqref{eq:Q5}, respectively.
 The results of the superfluid weight along the phase boundary for $\alpha=2T_0$ 
 at $T=0.35T_0$ are
 plotted as a function of $V$ in Fig.~\ref{fig:case5}(d), where the critical value of $V_c$ is shown with 
 a vertical broken line.  At $V=0$, $q$ is negative because of the odd-frequency Cooper pairs induced by the SOI.
 With increasing $V$, $|q_{\mathrm{odd}}|$ decreases and $q_{\mathrm{singlet}}$ increases. The spin-singlet 
 Cooper pairs compensate the negative superfluid weight due to odd-frequency pairs. 
 As a result, superconducting state becomes stable. The superfluid weight of the principal component $q_0$
 increases with increasing $V$ by reflecting the compensation. 
  By using Eqs.~\eqref{eq:Q4} and \eqref{eq:Q5}, it is possible to calculate the Zeeman field 
  at the compensation $V_{\mathrm{cross}}$ from 
\begin{align}  
  q_{\mathrm{odd}}(T, V_{\mathrm{cross}}) +q_{\mathrm{singlet}}(T, V_{\mathrm{cross}})=0.
\end{align}   
  The results are plotted on $H$-$T$ phase diagram in Fig.~\ref{fig:cross} in Appendix~\ref{secap:q5} for 
  three choices of $\alpha$. 
  The results show that $V_{\mathrm{cross}}$ increases almost linearly with $T$. 
To realize higher $T_c$, larger Zeeman fields are necessary. This correlation qualitatively explains 
a fact that the phase boundaries due to the second order transition in Fig.~\ref{fig:case5}(c) have 
almost linear line shape for all $\alpha$. 
  We conclude that the spin-singlet pairing correlation induced by the interplay 
between the SOI and the Zeeman field removes the instability due to the odd-frequency pairing correlation.
As a consequence, the superconducting phase is stable at high Zeeman fields.


\section{Properties of ZFIS}\label{sec:property}

\begin{figure}[t]
  \centering
  \includegraphics[width=\linewidth]{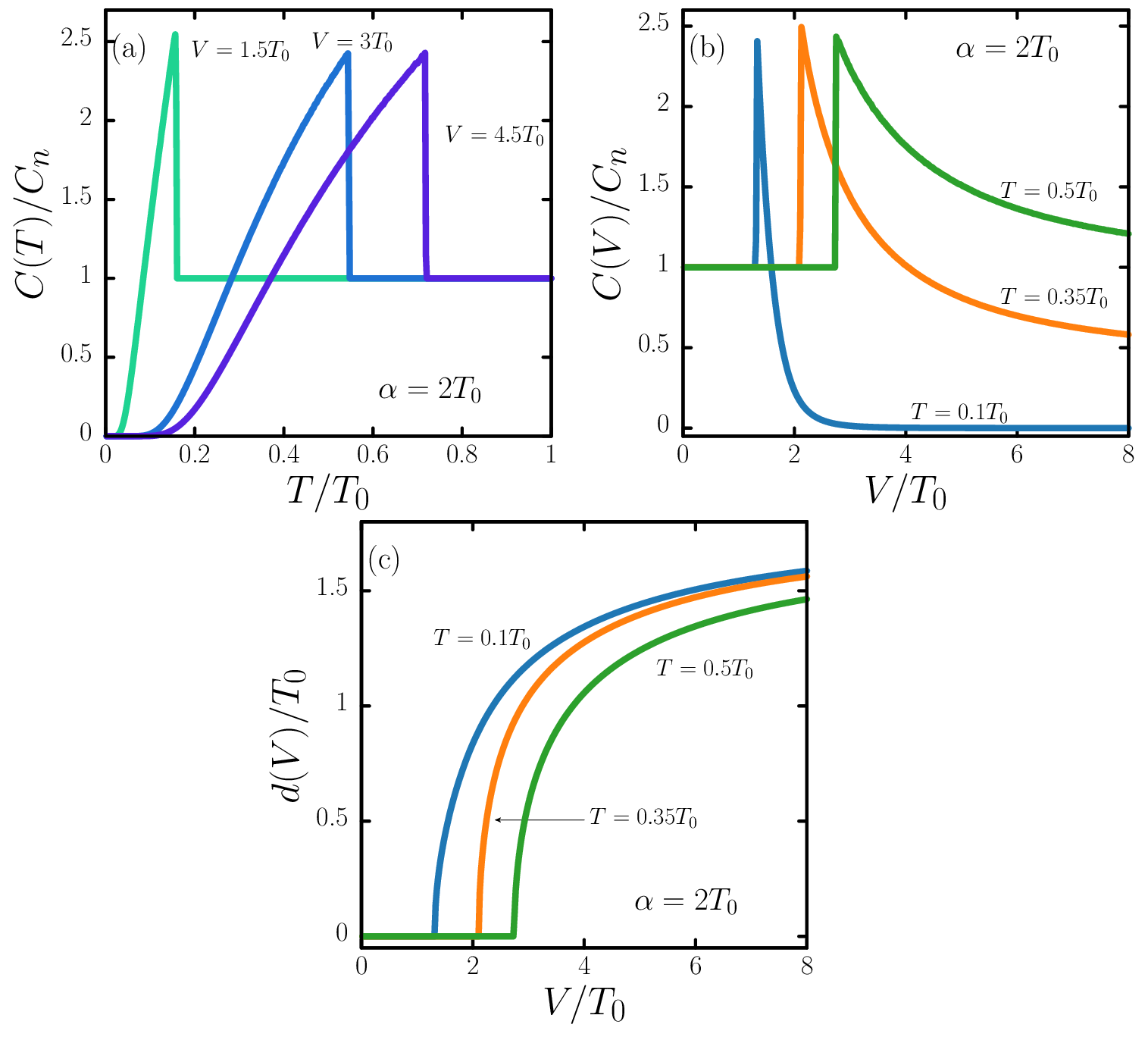}
  \caption{(a) The specific heat $C$ for Case 5.
  The dependence of the specific heat on temperature and that on Zeeman fields 
are shown in (a) and (b), respectively.  
  In (a), all three curves calculated for different Zeeman fields exhibit a discontinuous jump at the critical 
  temperature, characteristic of a second-order phase transition.
 The specific heat also indicates a jump at the critical Zeeman fields in (b).
Such singular behavior originates from the
 nonanalytic behavior of the pair potential at the critical Zeeman field as shown in (c).
  }
  \label{fig:specific_heat}
\end{figure}

In Sec.~\ref{sub5}, we discussed the phase diagram of Case 5 which indicates ZFIS.
Here we discuss properties of the superconducting phase by calculating 
the observables such as specific heat, magnetization, and magnetic susceptibility.
The specific heat calculated from
\begin{align}
  C=-T\frac{\partial^2\Omega_S}{\partial T^2},\label{eq:specific_heat}
\end{align}
is shown in Fig.~\ref{fig:specific_heat}(a), where 
$C_n$ is the specific heat of the normal state at $T_c$ and superconducting state is absent at $V=0$ 
because of large SOI $\alpha=2T_0 > V_{\mathrm{P}}$.
We set the Zeeman field to $V=1.5T_0$, $3T_0$ and $4.5T_0$.
All curves exhibit a discontinuous jump at the critical temperature, which indicates a second-order transition.
The specific heat is plotted as a function of Zeeman fields for several choices of 
temperatures in Fig.~\ref{fig:specific_heat}(b).
The results show a discontinuity of the specific heat at the critical Zeeman field $V_c$, which suggests 
 the rapid change in the pair potential there.
 In fact, the pair potential $d(V)$ plotted as a function of $V$ in
 Fig.~\ref{fig:specific_heat}(c) indicates that $d(V)$ starts to grow non-analytically at $V=V_c$.

\begin{figure}[t]
  \centering
  \includegraphics[width=\linewidth]{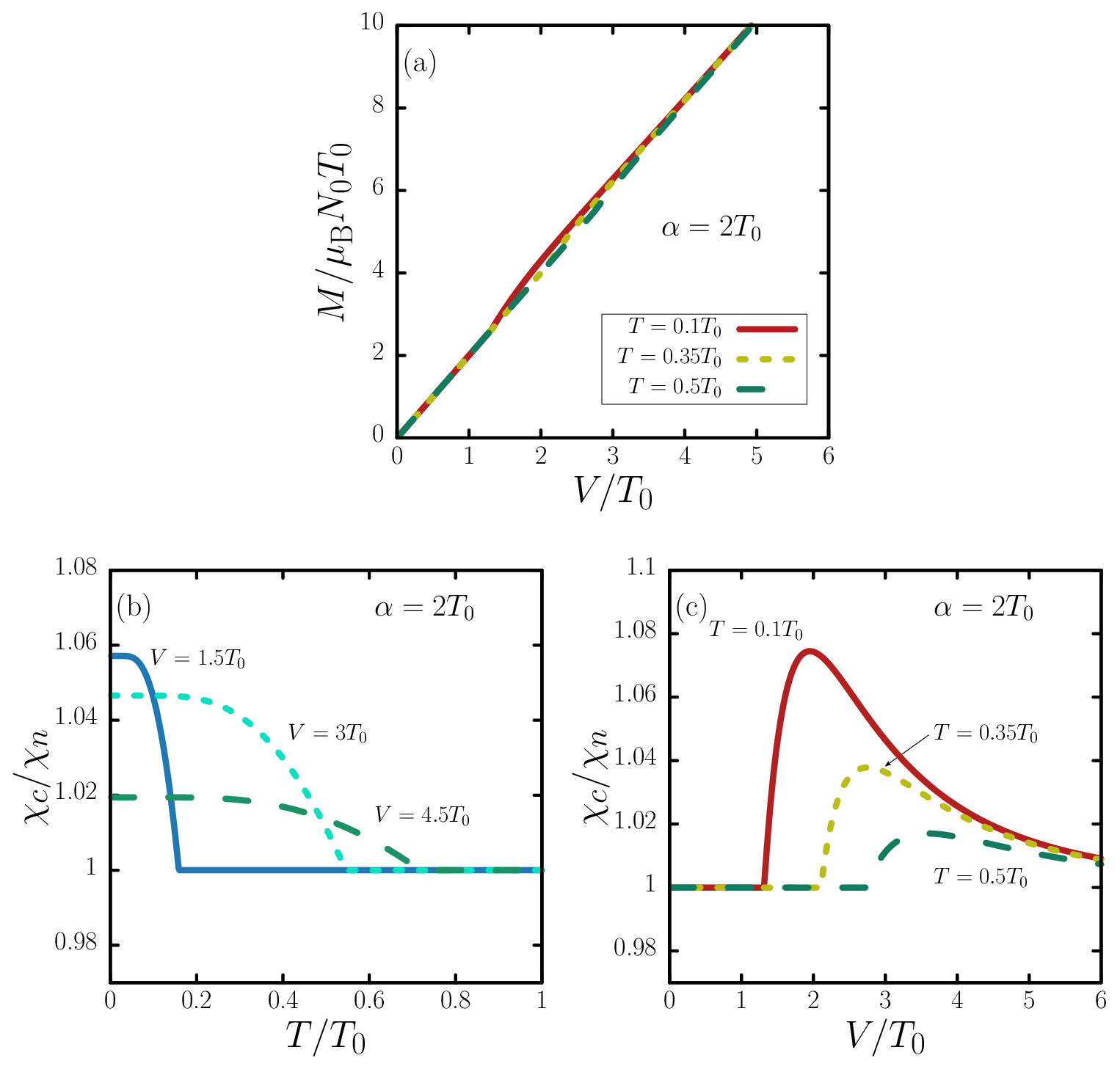}
  \caption{The magnetic properties for Case 5 at $\alpha=2T_0$.
The magnetization is plotted as a function of Zeeman field for several choices of temperature in (a).
 While the magnetization depends almost linearly on the Zeeman field, 
it slightly deviate the linear relation near the critical Zeeman field.
The deviation can be seen clearer in the magnetic susceptibility 
plotted as a function of $T$ in (b) and that plotted as a function of $V$ in (c). 
}
  \label{fig:magnetization_dos}
\end{figure}

Magnetic properties of superconducting states are studied
by measuring the magnetic moment and the magnetic susceptibility in experiments. 
The magnetization is defined as 
\begin{align}
  M&=-\mu_{\text{B}}\frac{\partial\Omega_S}{\partial V}\label{eq:magnetization}.
\end{align}
We supply the details of the derivation in Appendix~\ref{secap:chi_SS}.
The magnetization is plotted as Zeeman fields in Fig.~\ref{fig:magnetization_dos}(a) for several 
choices of temperature at $\alpha=2 T_0$. 
The results show the almost linear dependence of $M$ on Zeeman fields.
The deviation from the linear relationship can be seen around the critical Zeeman field.
Magnetic susceptibility enables us check such small structures.  
Here, we define three magnetic susceptibilities.
\begin{align} 
 \chi_c \equiv &\mu_{\text{B}}\frac{M}{V}, \label{eq:sus_c_def}\\
 \chi_d \equiv  &\mu_{\mathrm{B}} \frac{d M}{d V}, \label{eq:diff_mag_sus_def}\\
  \chi_{s}\equiv&\chi_n-\mu_{\text{B}}^2\frac{T}{4}\sum_{\bm{k}}\sum_{\omega_n}\text{Tr}\,\biggl[\check{\Gamma}\check{G}_5(\bm{k},\omega_n)\check{\Gamma}\check{G}_5(\bm{k},\omega_n)\biggr]
  \notag\\
  &\quad+\mu_{\text{B}}^2\frac{T}{4}\sum_{\bm{k}}\sum_{\omega_n}\text{Tr}\,\biggl[\check{\Gamma}\check{G}^N_5(\bm{k},\omega_n)\check{\Gamma}\check{G}^N_5(\bm{k},\omega_n)\biggr],\label{eq:spin_crore_def}
\end{align}
with
\begin{align}
  \check{{G}}_5(\bm{k},\omega_n)=&
  \begin{bmatrix}
    \hat{\mathcal{G}}_5&\hat{\mathcal{F}}_5\\
    -\undertilde{\hat{\mathcal{F}}}_5&-\undertilde{\hat{\mathcal{G}}}_5
  \end{bmatrix}_{(\bm{k},\omega_n)},\\
  \check{\Gamma}=&\frac{1}{V}
  \begin{bmatrix}
    \bm{V}\cdot\hat{\bm{\sigma}}\hat{\rho}_0&0\\
    0&-\bm{V}\cdot\hat{\bm{\sigma}}^*\hat{\rho}_0
  \end{bmatrix},
\end{align}
where $\check{\mathcal{G}}^N_5(\bm{k},\omega_n)$ is the Green's function of the normal state.
$\chi_c$ is calculated from the coefficient between $M$ and $H$, 
$\chi_d$ is the differential susceptibility, and 
$\chi_s$ is calculated from the spin-spin correlation function. 
In what follows, we focus on the Zeeman-field induced superconducting phase 
at $\alpha=2T_0$  in Fig.~\ref{fig:case5}(c) and show the three susceptibilities 
with the same parameters.  
In Fig.~\ref{fig:magnetization_dos}, $\chi_c$ is plotted as a function of temperature in (b) 
and as a function of $V$ in (c).
The results for $V<V_c$ indicates the Pauli's paramagnetism 
$\chi_{n}=2N_0\mu_{\text{B}}^2$ in (c).
The phase boundary in Fig.~\ref{fig:case5}(c) shows that almost linear relationship 
between $V_c$ and $T_c$.  The results in Fig.~\ref{fig:magnetization_dos}(b) and (c) 
show that  $\chi_c$ in superconducting phase is slightly larger than $\chi_n$
and saturates in the low-temperature region as shown in Fig.~\ref{fig:magnetization_dos}(b).
The deviation from $\chi_n$ is derived from the dependence of 
the pair potential $d$ on temperature in Fig.~\ref{fig:magnetization_dos}(b)
and that on $V$ in Fig.~\ref{fig:magnetization_dos}(c), respectively. 

\begin{figure}[t]
  \centering
  \includegraphics[width=\linewidth]{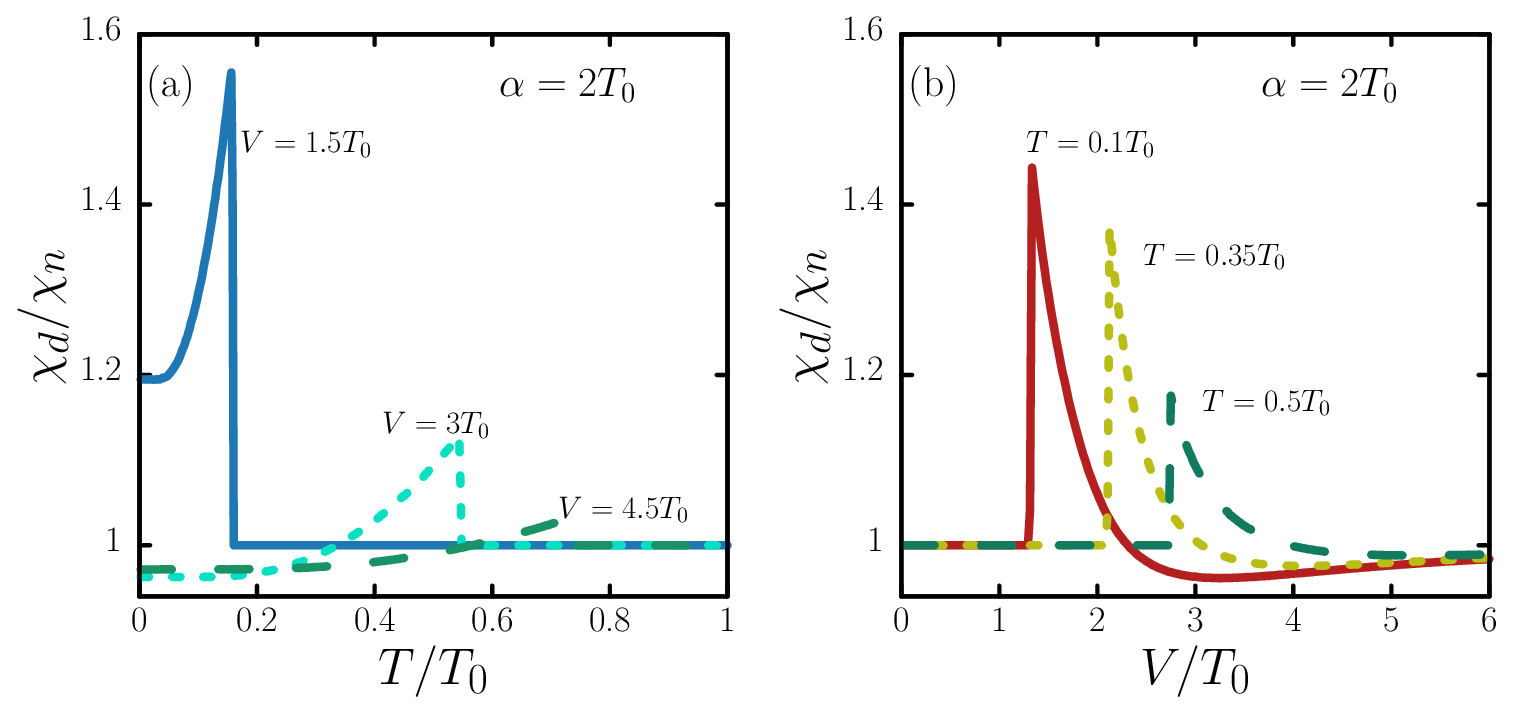}
  \caption{ The differential susceptibility $\chi_d$ in Eq.~\eqref{eq:diff_mag_sus_def} 
  at $\alpha=2.0T_0$, where $\chi_d$ versus $T$ is shown in (a) and 
  $\chi_d$ versus $V$ is shon in (b).
  These curves exhibit a discontinuous jump at the critical temperature, which originates 
  from the nonanalytic behabior of $d$ at the critical temperature and that at the critical Zeeman field. 
  }
  \label{fig:magnetic_sus}
\end{figure}

In  Fig.~\ref{fig:magnetic_sus}, the differential susceptibility $\chi_d$ is plotted as a 
function of $T$ in (a) and as a function of $V$ in (b).
  The deviation from $\chi_n$ is more remarkable because the pair potential 
  is nonanalytic around the critical temperature as shown in Fig.~\ref{fig:case5}(b) 
  and around the critical Zeeman field as shown in Fig.~\ref{fig:specific_heat}(c). 
As a consequence, $\chi_d$ changes discontinuously at the phase boundary and 
decreases toward the deep in superconducting phase.
The deviation from $\chi_n$ at zero temperature becomes negative for large $V$.

\begin{figure}[t]
  \centering
  \includegraphics[width=\linewidth]{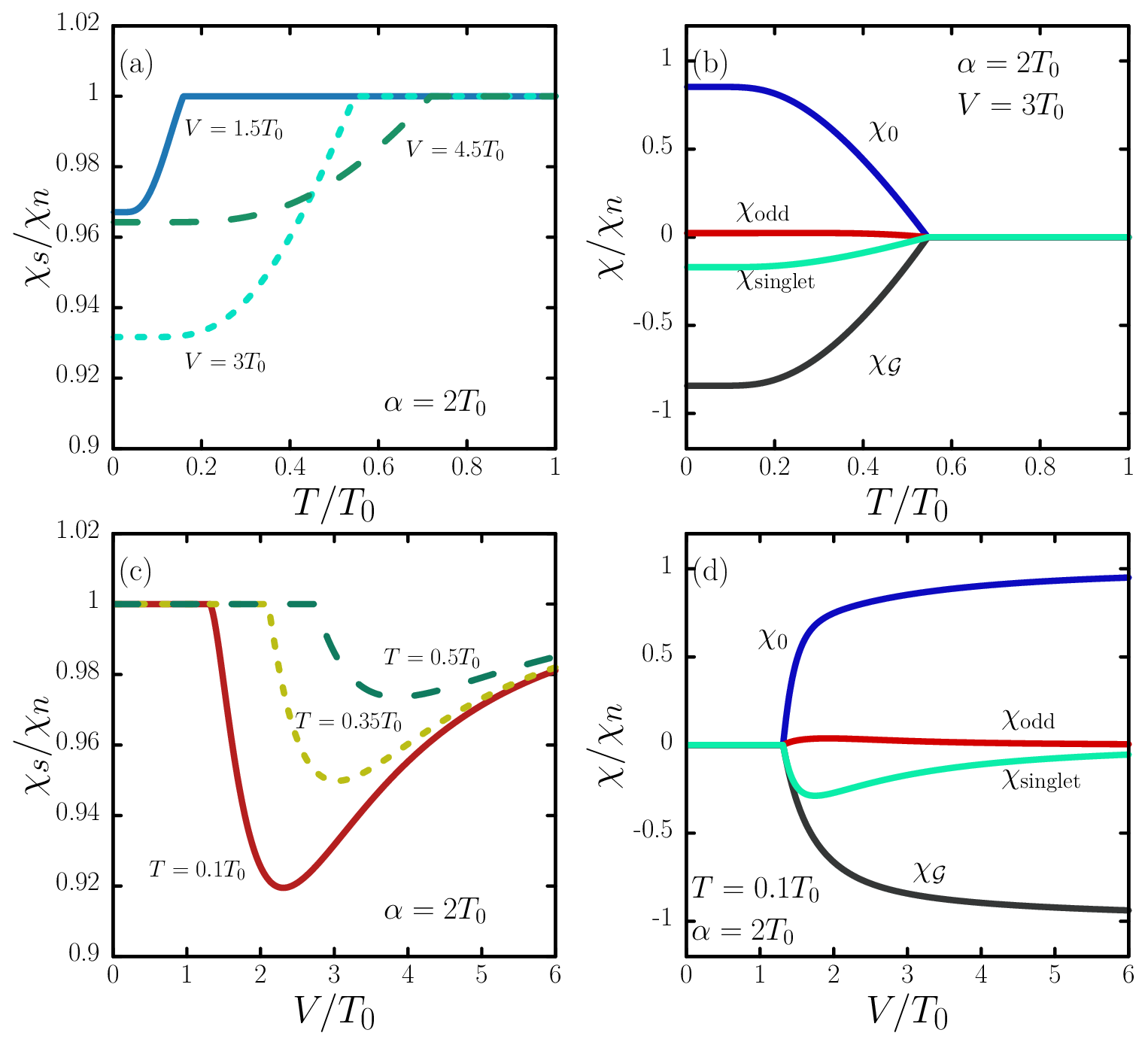}
  \caption{The susceptibility calculated from the spin-spin correlation function $\chi_{s}$ 
  in Eq.~\eqref {eq:spin_crore_def}. The amplitude of the SOI is fixed at $\alpha=2 T_0$.
  Figure (a) and (c) represents the dependence of $\chi_{s}$ on temperatures and that on Zeeman fields, 
  respectively.
  The correlation function is decomposed into four terms as shown in Eq.~\eqref{eq:spin_corre_separated}.
  They are plotted as a function of temperature at $V=3T_0$ in (b) and as a function of Zeeman fields 
  at $T=0.1T_0$ in (d).
The contribution of the normal Green's function
  $\chi_{\mathcal{G}}$ and that of the principal component $\chi_{0}$ almost cancel each other.
The susceptibility of odd-frequency component $\chi_{\text{odd}}$ is paramagnetic, whereas 
that of spin-singlet component $\chi_{\text{singlet}}$ is diamagnetic.
}
  \label{fig:spin_spin_corre_case5}
\end{figure}

The results of 
the spin-spin correlation function $\chi_s$ are shown as a function of $T$ in Fig.~\ref{fig:spin_spin_corre_case5}(a) and as a function of $V$ 
in Fig.~\ref{fig:spin_spin_corre_case5}(c). The spin-spin correlation decreases with decreasing temperature as shown in Fig.~\ref{fig:spin_spin_corre_case5}(a).
The degree of reduction is much smaller than $\chi_n$ and depens sensitively on $V$.   
The results in (c) show the reentrant behavior of $\chi_s$. 
The spin-spin correlation function can be decomposed into four terms
\begin{align}
  \chi_{s}=\chi_{\mathcal{G}}+\chi_{0}+\chi_{\text{odd}}+\chi_{\text{singlet}},\label{eq:spin_corre_separated}
\end{align}
as represented in Appendix~\ref{secap:chi_SS}.
The contribution of the normal Green's function is denoted by $\chi_{\mathcal{G}}$ 
and that of the principal component is given by $\chi_{0}$. 
$\chi_{\text{odd}}$ and $\chi_{\text{singlet}}$ are the susceptibility of odd-frequency 
pairing correlation and spin-singlet pairing correlation, respectively.
The results for $V=3 T_0$ are shown in Fig.~\ref{fig:spin_spin_corre_case5}(b) and those for $T=0.1T_0$ are shown in Fig.~\ref{fig:spin_spin_corre_case5}(d).
In both (b) and (d), the two components $\chi_{\mathcal{G}}$ and $\chi_{0}$ almost cancel 
each other. Therefore, the susceptibilites of induced pairing correlations $\chi_{\text{odd}}$ and $\chi_{\text{singlet}}$
dominate the structures in the superconducting phase. 
The results indicate that $\chi_{\text{odd}}$ is positive (paramagnetic)~\cite{higashitani:prl2013} and 
$\chi_{\text{singlet}}$ is negative (diamagnetic). Since $|\chi_{\text{singlet}}|\gg \chi_{\text{odd}}$, 
the spin-spin correlation function is smaller than $\chi_n$ as shown in Figs.~\ref{fig:spin_spin_corre_case5} (a) and (c). 
The reduction and the reentrance in $\chi_s$ in (a) and (c) 
 reflects the those of $\chi_{\text{singlet}}$. Thus,  induced spin-singlet component 
 governs the susceptibility $\chi_s$ calculated by the spin-spin correlation.

\section{conclusion}\label{sec:conclusion}

We have studied the stability of spin-triplet superconductivity in the presence of a Zeeman field  $\bm{V}$ and a 
spin-orbit interaction (SOI) $\bm{\alpha}$ by solving the Bogoliubov-de Gennes (BdG) Hamiltonian, where
electronic states in the normal state and spin-triplet order parameters $\bm{d}$  in the superconducting state
are described by using the two-valley model.
We calculate the anomalous Green's functions by solving the Gor'kov equation 
and the free-energy of the superconducting state.
The transition temperature $T_c$ under the magnetic potentials is calculated in two ways:  solving the gap equation 
and searching the minimum of the free-energy. This analysis enables us to distinguish the phase boundary 
due to a first-order transition and  that due to a second-order transition. To understand the stability or instability 
of superconducting state, we also analyze the superfluid weights calculated from the anomalous Green's function.
The resulting anomalous Green's function indicates that the magnetic potentials induce various subdominant 
paring correlations  in addition to the principal pairing correlation forming the pair potential. 
Although the induced pairing correlations are not linked directly to the pair potential, they stabilize or destabilize 
the superconducting state depending on their frequency symmetry. The even-frequency pairing correlation 
stabilizes the superconducting state, whereas the odd-frequency pairing correlation 
makes the superconducting state unstable. Such difference is well described by the sign of their superfluid weights. 
In our two-valley model, a $\bm{d}$ vector and an SOI $\bm{\alpha}$ are represented by the constant vectors 
independent of wave numbers $\bm{k}$. 
As a result, the analytic expressions of the gap equation and those of the superfluid weights enables us to 
understand the physics behind the phase diagram.

The stability of superconductivity is closely related to the relative orientation of three vectors in spin space:
 a Zeeman potential $\bm{V}$, a SOI $\bm{\alpha}$, and an order parameter of spin-triplet superconductivity $\bm{d}$.
A Zeeman field with $\bm{V}\perp\bm{d}$ and an SOI with $\bm{\alpha}\parallel\bm{d}$ do not affect the stability 
of the spin-triplet superconducting state.
However, a Zeeman field with $\bm{V}\parallel\bm{d}$ and an SOI with $\bm{\alpha}\perp\bm{d}$ make the superconducting state 
unstable. The odd-frequency pairing correlations induced by these magnetic potentials represent the instability because 
they decrease the superfluid weight.
When the three vectors are perpendicular to one another, we demonstrate the Zeeman-field induced superconductivity.
In the absence of a Zeeman field, the SOI induces the odd-frequency pairing correlation 
proportional to $\bm{\alpha}\times \bm{d}$ that suppresses superconductivity. 
In high Zeeman fields, 
even-frequency spin-singlet pairing correlation proportional to $\bm{\alpha}\times \bm{V} \cdot \bm{d}$ compensates 
the negative superfluid weight due to the odd-frequency pairing correlation.
As a result of the compensation, the amplitude of the principal pairing correlation increases with increasing Zeeman 
fields. The Jaccarino-Peter compensation has been accepted as a reasonable explanation for the Zeeman-field 
induced superconductivity.  The results in this paper provide an alternative explanation for the phenomenon. 
We also discuss the thermal and magnetic properties of Zeeman-field induced superconducting phase.

\section*{Acknowledgments}
T. S. is supported by JST SPRING, Grant Number JPMJSP2119.
S. I. is supported by a Grant-in-Aid for Early-Career Scientists (JSPS KAKENHI Grant No. JP24K17010).
Y. A. is supported by a Grant-in-Aid for Scientific Research (JSPS KAKENHI Grant No. JP26K0692).

\appendix
\section{Equality of the two contributions} \label{secap:qg_qf}
In this appendix, we show that $Q_{\mathcal{G}} = Q_{\mathcal{F}}$.
We describe the normal state Hamiltonian
by separating it into the kinetic energy and an arbitrary potential as
\begin{align}
  \check{H}_{\mathrm{N}} (\bm{k})
  =
  \xi_{\bm{k}} \check{1} + \check{V}_{\bm{k}} .
\end{align}
Here, $\check{\cdots}$ denotes the $N \times N$ matrix 
with $N$ being the number of degrees of freedom, such as spins, orbitals, and valleys.
The $N \times N$ identity matrix is represented by $\check{1}$.
Using the relation $\mathcal{G} \mathcal{G}^{-1} = \check{1}$ with
\begin{align}
  \mathcal{G} 
  =
  \begin{bmatrix}
    \check{\mathcal{G}} & \check{\mathcal{F}} \\
    - \undertilde{\check{F}} & - \undertilde{\mathcal{G}}
  \end{bmatrix} ,
\end{align}
we obtain
\begin{align}
  \frac{\partial \mathcal{G}}{\partial \xi_{\bm{k}}} \mathcal{G}^{-1}
  +
  \mathcal{G} \frac{\partial \mathcal{G}^{-1}}{\partial \xi_{\bm{k}}}
  =
  0 .
\end{align}
As a result, the derivative is expressed as
\begin{align}
  \frac{\partial \mathcal{G}}{\partial \xi_{\bm{k}}}
  &=
  \begin{bmatrix}
    \check{\mathcal{G}} \check{\mathcal{G}} + \check{\mathcal{F}} \undertilde{\check{\mathcal{F}}} &
    \check{\mathcal{G}} \check{\mathcal{F}} + \check{\mathcal{F}} \undertilde{\check{\mathcal{G}}} \\
    - \undertilde{\check{\mathcal{F}}} \check{\mathcal{G}} - \undertilde{\check{\mathcal{G}}} \undertilde{\check{\mathcal{F}}} &
    - \undertilde{\check{\mathcal{F}}} \check{\mathcal{F}} - \undertilde{\check{\mathcal{G}}} \undertilde{\check{\mathcal{G}}}
  \end{bmatrix} ,
\end{align}
where we assume that the potential $\check{V}_{\bm{k}}$
and the pair potential $\check{\Delta}_{\bm{k}}$ do not depend on $\xi_{\bm{k}}$,
which leads to 
\begin{align}
  \frac{\partial \mathcal{G}^{-1}}{\partial \xi_{\bm{k}}}
  &=
  \frac{\partial}{\partial \xi_{\bm{k}}}
  \begin{bmatrix}
    (i \omega_n - \xi_{\bm{k}})\check{1} - \check{V}_{\bm{k}} & - \check{\Delta}_{\bm{k}} \\
    \undertilde{\check{\Delta}}_{\bm{k}} & (i \omega_n + \xi_{\bm{k}})\check{1} + \undertilde{\check{V}}_{\bm{k}}
  \end{bmatrix}
  , \nonumber \\
  &=
  \begin{bmatrix}
    - \check{1} & \check{0} \\
    \check{0} & \check{1}
  \end{bmatrix} .
\end{align}
For the normal state, we obtain
\begin{align}
  \frac{\partial \mathcal{G}^{\mathrm{N}}}{\partial \xi_{\bm{k}}}
  =
  \begin{bmatrix}
    \check{\mathcal{G}}^{\mathrm{N}} \check{\mathcal{G}}^{\mathrm{N}} & \check{0} \\
    \check{0} & - \undertilde{\check{\mathcal{G}}}^{\mathrm{N}} \undertilde{\check{\mathcal{G}}}^{\mathrm{N}}
  \end{bmatrix},
\end{align}
where we use the Green's function in the normal state given by
\begin{align}
  \check{\mathcal{G}}^{\mathrm{N}}
  =
  \left[ (i \omega_n - \xi_{\bm{k}}) \check{1} - \check{V}_{\bm{k}} \right]^{-1} .
\end{align}
Using these results, we find the relation
\begin{align}
  \check{\mathcal{G}} \check{\mathcal{G}} - \check{\mathcal{G}}^{\mathrm{N}} \check{\mathcal{G}}^{\mathrm{N}}
  =
  \frac{\partial}{\partial \xi_{\bm{k}}} \left( \check{\mathcal{G}} - \check{\mathcal{G}}^{\mathrm{N}} \right)
  - 
  \check{\mathcal{F}} \undertilde{\check{\mathcal{F}}} .
\end{align}
As a result, the contribution to the superfluid weight from the normal Green's function becomes
\begin{align}
  Q_{\mathcal{G}}
  &=
  \frac{1}{2} T \sum_{\omega_n} \int_{- \infty}^{\infty} d \xi_{\bm{k}} \, \mathrm{Tr} \left[
    \frac{\partial}{\partial \xi_{\bm{k}}} 
    \left( \check{\mathcal{G}} - \check{\mathcal{G}}^{\mathrm{N}} \right)
    - 
    \check{\mathcal{F}} \undertilde{\check{\mathcal{F}}}
  \right]
  , \nonumber \\
  &=
  \frac{1}{2} T \sum_{\omega_n} \int_{- \infty}^{\infty} d \xi_{\bm{k}} \, \mathrm{Tr} \left[
    - \check{\mathcal{F}} \undertilde{\check{\mathcal{F}}}
  \right]
  , \\
  &= 
  Q_{\mathcal{F}} .
\end{align}

\section{Superfluid weight}\label{secap:qf}

By substituting the anomalous Green's function 
in Eq.~\eqref{eq:f} into Eq.~\eqref{eq:qf_def}, the superfluid weight is decomposed into
\begin{widetext}
\begin{align}
  Q_{\mathcal{F}}&=Q_0
  +Q_{\bm{V}\parallel\bm{d}}+Q_{\bm{\alpha}\perp\bm{d}}
  +Q_{\bm{V}\parallel\bm{\alpha}}
  +Q_{\bm{V}\perp\bm{\alpha}\perp\bm{d}},\\
  Q_0&=\frac{1}{2}T\sum_{\omega_n}\int d\xi\,
  \mathrm{Tr}\bigl[\frac{1}{\check{Z}^{2}}
  \{(\omega_n^{2}+\hat{\xi}_{\bm{k}}^{2}+d^{2}-\alpha^{2}+V^{2})\bm{d}
  +2(\bm{\alpha}\cdot\bm{d})\bm{d}-2(\bm{V}\cdot\bm{d})\bm{d}\}^{2}\bigr], \nonumber \\
%
  &+ \frac{1}{2}T\sum_{\omega_n}\int d\xi\,
  \mathrm{Tr}\bigl[\frac{1}{\check{Z}^{2}}
  4\hat{\xi}_{\bm{k}}^{2}(\bm{V}\times\bm{d})^{2}\bigr] 
%
  +\frac{1}{2}T\sum_{\omega_n}\int d\xi\,
  \mathrm{Tr}\bigl[\frac{1}{\check{Z}^{2}}
  4\hat{\xi}_{\bm{k}}^{2}(\bm{\alpha}\cdot\bm{d})^{2}\bigr],\label{eq:Q0}\\
  Q_{\bm{V}\parallel\bm{d}}&=-\frac{1}{2}T\sum_{\omega_n}\int d\xi\,
  \mathrm{Tr}\bigl[\frac{1}{\check{Z}^{2}}
  4\omega_n^{2}(\bm{V}\cdot\bm{d})^{2}\bigr],\label{eq:Q3}\\
  Q_{\bm{\alpha}\perp\bm{d}}&=-\frac{1}{2}T\sum_{\omega_n}\int d\xi\,
  \mathrm{Tr}\bigl[\frac{1}{\check{Z}^{2}}
  4\omega_n^{2}(\bm{\alpha}\times\bm{d})^{2}\bigr],\label{eq:Q4}\\
  Q_{\bm{V}\perp\bm{\alpha}\perp\bm{d}}&=\frac{1}{2}T\sum_{\omega_n}\int d\xi\,
  \mathrm{Tr}\bigl[\frac{1}{\check{Z}^{2}}
  4(\bm{V}\times\bm{\alpha}\cdot\bm{d})^{2}\bigr],\label{eq:Q5}\\
  Q_{\bm{V}\parallel\bm{\alpha}}&=\frac{1}{2} T\sum_{\omega_n}\int d\xi\,
  \mathrm{Tr}\bigl[\frac{1}{\check{Z}^{2}}
  8i\omega_n\hat{\xi}_{\bm{k}}(\bm{V}\cdot\bm{\alpha})d^{2}\hat{\rho}_z\bigr]\label{eq:Q6}.
\end{align}
As we discussed in Sec.~\ref{sec:results} A, 
the contributions of $Q_0$, $Q_{\bm{V}\perp\bm{d}}$ and $Q_{\bm{\alpha}\parallel\bm{d}}$ 
to the stability of superconducting states are inseparable.
The superfluid weights due to odd-frequency pairing correlations 
 $Q_{\bm{V}\parallel\bm{d}}$ and $Q_{\bm{\alpha}\perp\bm{d}}$ are negative.
 Eq.~\eqref{eq:Q6} originates from cross terms between the Cooper pairs which proportional to $\omega_n$ and $\hat{\xi}_{\bm{k}}$.
This term also decreases $Q_{\mathcal{F}}$.
Indeed, $Q_{\bm{V}\parallel\bm{\alpha}}$ satisfies
\begin{align}
  Q_{\bm{V}\parallel\bm{\alpha}}&=\frac{1}{2}T\sum_{\omega_n}\int d\xi\,
  \mathrm{Tr}\bigl[\frac{1}{\check{Z}^{2}}
  8i\omega_n\hat{\xi}_{\bm{k}}(\bm{V}\cdot\bm{\alpha})d^{2}\hat{\rho}_z\bigr]\notag\\
  &=-nT\sum_{\omega_n>0}\int d\xi\,
  \mathrm{Tr}\bigl[\frac{1}{(\check{Z}\check{Z}^{*})^{2}}
  64\omega_n^{2}\hat{\xi}_{\bm{k}}^{2}(\bm{V}\cdot\bm{\alpha})^{2}d^{2}\bigr]\notag.
\end{align}

\section{Superfluid weights in case 5} \label{secap:q5}

The superfluid weights defined in Eq.~\eqref{eq:sqdef} are calculated for case 5,
\begin{align}
q_0(T, V) =&  T\sum_{\omega_n>0} \frac{\pi}{\omega_n^3 D_0^3} \left[
(2\omega_n^2+ 2V^2 -\alpha^2) D_0^2 - \alpha^2(4 \omega_n^2 - \alpha^2) D_0 + 4 \alpha^4 \omega_n^2 
 \right],\\
q_{\mathrm{odd}}(T, V) =&  -  T\sum_{\omega_n>0} \frac{\pi}{\omega_n^3 D_0^3}  \alpha^2 \omega_n^2 (4\omega_n^2 + D_0), \\
q_{\mathrm{singlet}}(T, V) =&   T\sum_{\omega_n>0} \frac{\pi}{\omega_n^3 D_0^3}  \alpha^2 V^2 (4\omega_n^2 + D_0), \\
D_0 =& \omega_n^2 + \alpha^2+V^2,
\end{align}
where $q_{\mathrm{odd}}$ and $q_{\mathrm{singlet}}$ are calculated from Eq.~\ref{eq:Q4} and \eqref{eq:Q5}, respectively.
The cancelation of the negative superfluid weight  by the spin-singlet pairs is represented by
\begin{align}
q_{\mathrm{odd}}(T, V_{\mathrm{cross}})  + q_{\mathrm{singlet}}(T, V_{\mathrm{cross}})  =0.
\end{align}
 The solution of the equation $V_{\mathrm{cross}}$ is plotted as a function of temperatures in Fig.~\ref{fig:cross}
 for three choices of $\alpha$. The results show the almost linear relationship insensitive to $\alpha$.

\begin{figure}[t]
  \centering
  \includegraphics[width=0.4\linewidth]{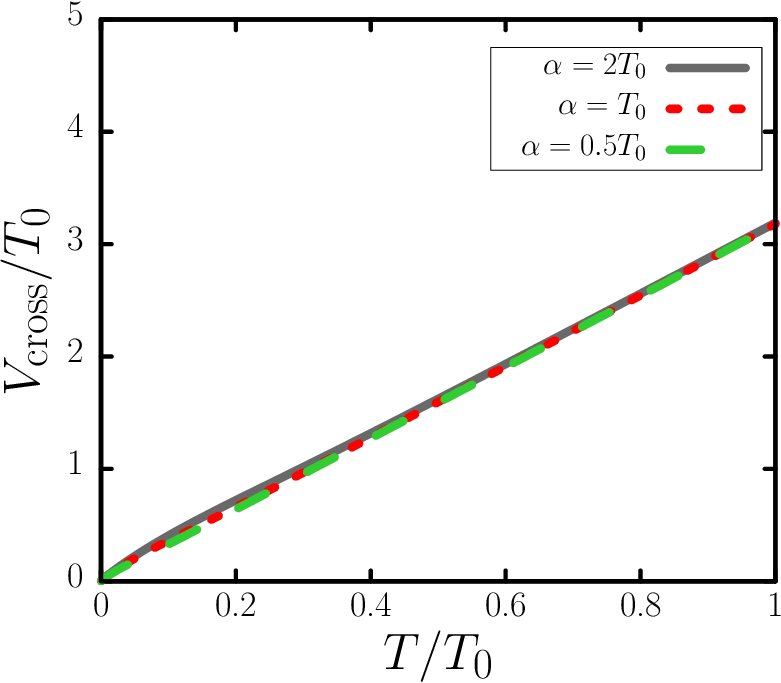}
  \caption{$V_{\mathrm{cross}}$ is the Zeeman fiels at which 
  the spin-singlet correlation cancels the negative superfluid weight due to odd-frequency pairing 
  correlation as shown in Fig.~\ref{fig:case5}(d).
  }
  \label{fig:cross}
\end{figure}

\section{Calculation of the magnetic properties of ZFIS}\label{secap:chi_SS}

The derivative with respect to $V$ that appears in the expression for the magnetization in Eq.~\eqref{eq:magnetization} is decomposed as 
\begin{align}
  M&=-\mu_{\text{B}}\frac{\partial \Omega_S}{\partial V}
  =-\mu_{\text{B}}\left(\frac{\partial \Omega_S}{\partial V}\right)_{d}
  -\mu_{\text{B}}\left(\frac{\partial \Omega_S}{\partial d}\right)_{V}
  \frac{\partial d}{\partial V},
\end{align}
where $(\partial X/\partial x)_y$ denotes the partial derivative of $X$ with respect to $x$ at constant $y$.
At equilibrium, the second term vanishes because $(\partial \Omega_S/\partial d)_V=0$ corresponds to the gap equation.
Thus, the magnetization is calculated as 
\begin{align}
  M&=-\mu_{\text{B}}\left(\frac{\partial \Omega_S}{\partial V}\right)_{d}\notag\\
  &=\frac{\mu_{\text{B}}N_0}{4}\sum_{\sigma,\rho=\pm1}\int d\xi\,\tanh\left(\frac{E_{\sigma,\rho,\bm{k}}}{2T}\right)\frac{\partial E_{\sigma,\rho,\bm{k}}}{\partial V}\notag\\
  &=\frac{\mu_{\text{B}}N_0}{4}\sum_{\sigma,\rho=\pm1}\int d\xi\,\tanh\left(\frac{E_{\sigma,\rho,\bm{k}}}{2T}\right)\frac{1+\sigma\xi^2_{\rho}/\sqrt{(\alpha^2+V^2)\xi^2_{\rho}+d^2\alpha^2}}{E_{\sigma,\rho,\bm{k}}}V,
\end{align}
where $E_{\sigma,\rho,\bm{k}}$ is defined in Eq.~\eqref{eq:eig_5}.

The spin-spin correlation function defined in Eq.~\eqref{eq:spin_crore_def} is decomposed into 
the four contributions
\begin{align}
  \chi_{s}&=\chi_{\mathcal{G}}+\chi_{0}+\chi_{\text{odd}}+\chi_{\text{singlet}}\\
  \chi_{\mathcal{G}}
  &=
  -2N_0T\sum_{\omega}\int d\xi\,\frac{1}{Z_5^2}
  \biggl[
  \xi^6+(\omega_n^2+2d^2-V^2-3\alpha^2)\xi^4,\\
  &\quad\quad
  +(-\omega_n^4+2(\alpha^2-5V^2)\omega_n^2+(\alpha^2+V^2-d^2)^2+2(\alpha^2+V^2)(\alpha^2-V^2-d^2))\xi^2\notag\\
  &\quad\quad+
  (V^2-\omega_n^2)(\omega_n^2+d^2+\alpha^2+V^2)^2-\alpha^2(\omega_n^2-d^2+\alpha^2+V^2)^2
  \biggr],\\
  \chi_{0}&=
  2N_0T\sum_{\omega}\int d\xi\,\frac{1}{Z_5^2}
  \biggl[(\omega_n^2+\xi^2+d^2-\alpha^2+V^2)^2+4\xi^2V^2\biggr]d^2,\\
  \chi_{\text{odd}}&=
  2N_0T\sum_{\omega}\int d\xi\,\frac{1}{Z_5^2}4\omega_n^2\alpha^2d^2, \\
  \chi_{\text{singlet}}&=-2N_0T\sum_{\omega}\int d\xi\,\frac{1}{Z_5^2}4V^2\alpha^2d^2,
\end{align}
where $Z_5$ is shown in Eq.~\eqref{eq:Z_5}.
The normal Green's function gives $\chi_{\mathcal{G}}$ and the principal component on the
anomalous Green's function gives $\chi_{0}$. 
The susceptibility of additional spin-singlet pairs $\chi_{\text{singlet}}$ is diamagnetic, 
whereas that of odd-frequency pairs $\chi_{\text{odd}}$ is paramagnetic.
The anomalous Green's function components $\chi_{0}$, $\chi_{\text{odd}}$ and $\chi_{\text{singlet}}$ have the same functional form as those of the superfluid density.

\end{widetext}

\bibliography{list_2026}

\end{document}